\documentclass[letterpaper,twocolumn,10pt]{article}

\usepackage{hegroup}
\usepackage{tikz}
\usepackage{amsfonts}
\usepackage{xspace} 
\usepackage{amsmath}
\usepackage{mathrsfs}
\usepackage{amssymb}
\usepackage{amsmath}
\usepackage{amsthm}
\usepackage{multirow}
\usepackage{makecell}
\usepackage[sort&compress,numbers]{natbib}
\usepackage{caption}
\usepackage{subcaption}
\usepackage[inkscapelatex=false]{svg}
\usepackage{float}
\usepackage{graphicx}
\usepackage{booktabs}
\usepackage{balance}
\usepackage{xcolor}
\usepackage{tcolorbox}
\usepackage{colortbl}
\usepackage{hyperref}
\usepackage{url}
\usepackage{array}
\usepackage{diagbox}
\usepackage[utf8]{inputenc}
\usepackage{url}
\usepackage{cleveref}
\usepackage{booktabs}
\usepackage{amssymb}
\usepackage{bbding}
\usepackage{pifont}
\usepackage{wasysym}
\usepackage{utfsym}
\usepackage[misc]{ifsym}
\usepackage{fontawesome}
\usepackage{dutchcal}
\usepackage{svg}
\usepackage{scalerel,xparse}
\NewDocumentCommand\emojismile{}{
    \scalerel*{
        \includegraphics{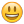}
    }{X}
}
\usepackage{textcomp}  
\usepackage{scalerel}  
\hypersetup{
  colorlinks,
  linkcolor={blue!70!green},
  citecolor={green!70!blue},
  urlcolor={orange!70!red}
}

\usepackage[hang,flushmargin]{footmisc}

\newcommand{\mypara}[1]{\noindent{\bf {#1}.}\xspace}

\begin{document}

\date{}
\pagestyle{plain} 

\title{On Evaluating The Performance of Watermarked Machine-Generated Texts Under Adversarial Attacks}

\author{
Zesen Liu\textsuperscript{1, 2}  \ \ \ 
Tianshuo Cong\textsuperscript{3}  \ \ \ 
Xinlei He\textsuperscript{1}\thanks{Corresponding author (\href{mailto:xinleihe@hkust-gz.edu.cn}{xinleihe@hkust-gz.edu.cn}).} \ \ \ 
Qi Li\textsuperscript{3} \ \ \ 
\\
\\
\textsuperscript{1}\textit{Hong Kong University of Science and Technology (Guangzhou)} \ \ \ 
\\
\textsuperscript{2}\textit{Xidian University} \ \ \
\textsuperscript{3}\textit{Tsinghua University} \ \ \ 
}

\maketitle

\begin{abstract}

Large Language Models (LLMs) excel in various applications, including text generation and complex tasks.
However, the misuse of LLMs raises concerns about the authenticity and ethical implications of the content they produce, such as deepfake news, academic fraud, and copyright infringement.
Watermarking techniques, which embed identifiable markers in machine-generated text, offer a promising solution to these issues by allowing for content verification and origin tracing.
Unfortunately, the robustness of current LLM watermarking schemes under potential watermark removal attacks has not been comprehensively explored.

In this paper, to fill this gap, we first systematically comb the mainstream watermarking schemes and removal attacks on machine-generated texts, and then we categorize them into pre-text (before text generation) and post-text (after text generation) classes so that we can conduct diversified analyses.
In our experiments, we evaluate eight watermarks (five pre-text, three post-text) and twelve attacks (two pre-text, ten post-text) across 87 scenarios.
Evaluation results indicate that 
(1) KGW and Exponential watermarks offer high text quality and watermark retention but remain vulnerable to most attacks;
(2) Post-text attacks are found to be more efficient and practical than pre-text attacks;
(3) Pre-text watermarks are generally more imperceptible, as they do not alter text readability, unlike post-text watermarks;
(4) Additionally, combined attack methods can significantly increase effectiveness, highlighting the need for more robust watermarking solutions.
Our study underscores the vulnerabilities of current techniques and the necessity for developing more resilient schemes.

\end{abstract}

\section{Introduction}

Large Language Models (LLMs) are rapidly advancing in capability, demonstrating remarkable proficiency across a wide range of applications.
From generating coherent and contextually appropriate text to assisting in complex tasks such as code generation~\cite{chen2021evaluating,guo2023longcoder,lu2022reacc,wang2023codet5,wu2024repoformer}, medical diagnosis~\cite{jin2024healthllm,singhal2022large,gao2023leveraging}, and content creation~\cite{goldstein2023generative,Lim2021HackingHumans,OpenAI2023GPT4SystemCard}, LLMs are revolutionizing various domains.
Their integration into these fields highlights the transformative potential of artificial intelligence, providing enhanced productivity and innovative solutions to longstanding challenges.

Despite their powerful capabilities, the rise of LLMs has also sparked significant concerns regarding the authenticity and ethical implications of the content they generate.
Issues such as the creation of deepfake texts~\cite{uchendu2023does,mitra2024world,pu2022deepfake}, automated homework and programming assignments~\cite{autod}, and the spread of misinformation~\cite{pan2023risk,10.1145/3544548.3581318} pose serious risks to the integrity of information and trust in digital communications.
The ability of LLMs to produce highly realistic and human-like text exacerbates these concerns, making it increasingly difficult to distinguish between human and machine-generated content.

To address these challenges, watermark techniques~\cite{article_watermark,George22} have emerged as promising solutions.
By embedding identifiable markers within the machine-generated text, these techniques aim to provide a reliable means of tracing the origin of the text and verifying its authenticity.
This approach offers a potential safeguard against the misuse of LLMs, helping to preserve the credibility of information and enhance accountability in content creation.

However, the robustness of these watermarking schemes remains questionable.
Adversaries may develop methods to circumvent or remove watermarks, undermining their effectiveness and potentially rendering them unreliable.
The resilience of watermarking techniques against various forms of manipulation and obfuscation is critical to their success and necessitates rigorous evaluation to ensure their practicality in real-world scenarios.

\mypara{Our work}
In this paper, we fill the gap in the literature by systematically categorizing watermarks and attacks into pre-text and post-text classes, where the former involves watermark injection or disruption before text generation, and the latter does so after text generation.
We consider eight watermarks (five pre-text and three post-text) and twelve attacks (two pre-text and ten post-text), resulting in a total of 87 possible scenarios (see \Cref{tab: attack and watermark}).
Our evaluation shows that KGW~\cite{kirchenbauer2023watermark} and Exponential~\cite{aaronson2022watermarking} are the two best watermarks by providing good text quality and relatively high watermark rates after different attacks compared to other watermarks.
For instance, KGW and Exponential achieve over 0.5 robustness score while Linguistic~\cite{yang2023watermarking} watermarks only reach around 0.3 robustness score (see \Cref{tab: robustness of all} for more details).
However, they are still not robust against most of the attacks.
For example, KGW only reaches a 0.0349 watermark rate under Paraphrase attacks.
This suggests that existing watermarks are still vulnerable to various attacks.
Regarding efficiency, both KGW and Exponential are relatively efficient in watermark injection and detection.
In terms of attacks, we find that post-text attacks are generally more efficient than pre-text attacks since they do not require modification of the model's weights, making them more practical threats to current watermarks.
Concerning imperceptibility, pre-text watermarks usually perform better as they are hidden in the token distributions and do not affect text readability, whereas post-text watermarks involve adding, removing, or replacing tokens, making them easier to detect.

From the attack perspective, we find that most post-text attacks are much more efficient than the pre-text attacks as they do not require modification of the model's weight, which makes them more practical threats to current watermarks.
We also find that adversaries can significantly enhance their attack effectiveness by combining different attack methods.
For instance, KGW's watermark rate decreases to 0.2365 when applying the Synonym attack first and the Modify attack later (see \Cref{fig: results of watermark text} for more details).
This underscores the need to develop more robust watermarking solutions to withstand diverse attacks.
Our code and data will be made publicly available.

In this paper, we make the following contributions.

\begin{itemize}
    \item We perform the most comprehensive (thus far) evaluation of the performance of watermarked machine-generated text under adversarial attacks.
    \item We develop a uniform framework for constructing watermarks/attacks, defining metrics from different aspects, and evaluating the performance.
    \item Our evaluations indicate that current watermarks are highly susceptible to adversarial attacks, emphasizing the necessity for more robust watermarking schemes.
\end{itemize}

\section{Background}

\subsection{Large Language Models~(LLMs)}

In recent years, Large Language Models (LLMs) have achieved remarkable success in the field of Natural Language Processing (NLP).
These models, trained on extensive datasets, have demonstrated unparalleled language understanding and generation capabilities.
They have revolutionized various applications, from conversational agents like OpenAI's ChatGPT and GPT-4 to advanced systems such as Google's Gemini.
LLMs have not only excelled in traditional language tasks but have also proven instrumental in solving complex, real-world problems, greatly enhancing human productivity and interaction with technology. 

Following~\cite{naseh2023stealing}, we model the generation process of LLMs into two steps: probability prediction and token selection.
First, given a sequence of tokens $x_{1:n}$ (e.g., the input prompt) where $x_i \in \{1,...,|V|\}$ is a token and $|V|$ is the size of vocabulary $V$, LLM calculates the conditional probability of the next token as $Pr(x_{n+1}|x_{1:n})$.
Then, LLM uses a specific sampling strategy to select the word from the vocabulary as the final token, e.g., Greedy Sampling confirms the next token by $x_{n+1} = \arg \max_{x} Pr(x|x_{1:n})$.
Naturally, LLM generates a $l$ length text through computing $Pr(x_{1:n+l}) = \prod_{i=1}^{n+l}Pr(x_i|x_{1:i-1})$.

\subsection{Preliminary of Watermarks}

\begin{figure*}
\centering
\includegraphics[width=1\linewidth]{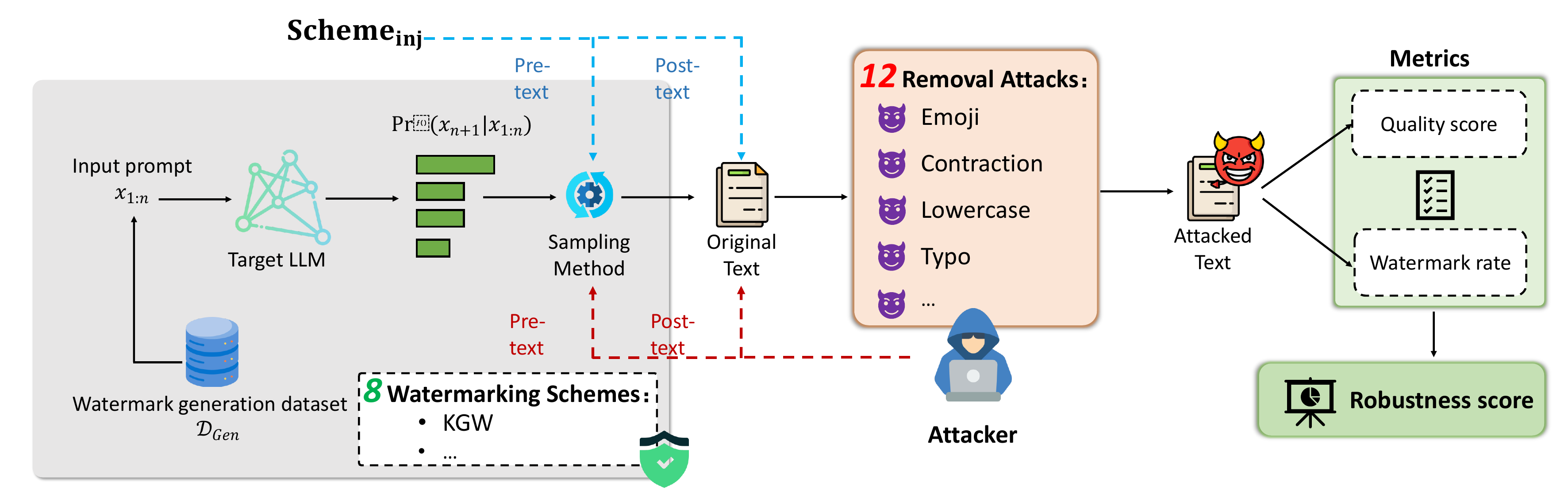}
\caption{Workflow of our evaluation framework. We consider eight mainstream LLM watermarking schemes and twelve watermark removal attacks, resulting in a total of 87 possible scenarios.}
\label{fig:overview}
\end{figure*}

We first classify the watermarking schemes from different perspectives.
For the access needed to the model's internal parameters, we classify schemes into black-box and white-box.
Meanwhile, as~\Cref{fig:overview} shows, according to the watermarking scheme with respect to the time involved in the text generation process, we classify them into pre-text schemes and post-text schemes.
Note that the same classification rules can be applied to watermark removal attacks, that is, we will discuss white- and black-box removal attacks, as well as pre-text and post-text attacks.
Nevertheless, regardless of the classification, an LLM watermarking scheme typically contains two processes: \textit{watermark injection} and \textit{watermark detection}.

\mypara{Black-box \& White-box} 
A watermarking scheme is classified as \textit{white-box} if it relies on the model’s internal parameters during watermark injection. 
Conversely, a scheme is \textit{black-box} if it does not access model parameters. 
Notably, schemes accessing the last hidden state of an LLM are also considered black-box.

\mypara{Pre-text \& Post-text} 
We further categorize a watermarking scheme based on the time it is conducted, labeling it as either pre-text or post-text.
If it is conducted before or during the text generation phase, we classify it as a pre-text watermarking scheme.
Conversely, we regard it as a post-text watermarking scheme.

\mypara{Watermark injection}
During the watermark injection process, a multitude of distinct scheme-related hyperparameters are necessitated:
1) The pre-text watermarking schemes~\cite{kirchenbauer2023watermark,kuditipudi2023robust,aaronson2022watermarking,cryptoeprint:2023/763,zhao2023provable,lee2024wrote, lu2024entropybased,liu2024unforgeable,liu2023watermarking,he2024watermarks} usually requires a secret key and a pseudo-random number generator;
2) For the post-text watermarking schemes~\cite{article,sato2023embarrassingly,UniSpaCh,zhang2023remark,yang2023watermarking}, only a watermark signal~(e.g. the specific whitespace \texttt{$\setminus$u2004} or \texttt{$\setminus$u2006}) is needed.
In this paper, given a watermarking scheme, we denote all hyperparameters it used as $\kappa$.
Therefore, given input $x$, the watermark injection process can be defined as:
\begin{equation}
    y = \mathsf{Scheme}_{\rm inj}(x,\kappa; \mathcal{M}),
\end{equation}
where $\mathcal{M}$ is the target model that the scheme aims to protect and $y$ is the watermarked text.

\mypara{Watermark detection}\label{iniwatermark}
As the inverse process of the injection process, the detection process is defined as:
\begin{equation}
\{\textit{Ture}, \textit{False}\} = \mathsf{Scheme}_{\rm dec}(y,\kappa; \mathcal{M}),
\end{equation}
where \textit{True} stands for that we can extract the watermark information from model $\mathcal{M}$ by using the same $\kappa$.

\section{Watermarks}

First of all, we clarify that we focus on the watermarking schemes which are based on the generated texts in this paper. Compared with the watermarking schemes which need to embed the watermarks by updating the model parameters~\cite{peng-etal-2023-copying,li2023watermarking}, these text-based schemes can be adapted to any causal language models.
Meanwhile, we consider two types of watermarks, i.e., pre-text watermarks and post-text watermarks, where the former injects the watermark before or during text generation and the latter injects the watermark after the text generation.

The methods of pre-text watermarks can be divided into two categories based on the process used for generating text with an LLM~\cite{liu2024survey}.
For instance, this includes modifying the logits~\cite{kirchenbauer2023watermark,zhao2023provable,liu2024unforgeable,lee2024wrote,lu2024entropybased,he2024watermarks,liu2023watermarking} and modifying the token sampling strategy during the inference phase~\cite{kuditipudi2023robust,aaronson2022watermarking,cryptoeprint:2023/763}.
The methods of post-text watermark can be divided into four categories based on the granularity of text modification~\cite{liu2024survey}, which are format-based watermark~\cite{UniSpaCh,sato2023embarrassingly}, lexical-based watermark~\cite{yang2023watermarking,munyer2024deeptextmark}, synaptic-based watermark~\cite{article_synatic,article}, and generation-based watermark~\cite{zhang2023remark,abdelnabi2021adversarial}.
To comprehensively discuss the robustness of various types of watermarking schemes, we discuss eight LLM watermarking schemes in this paper, which are summarized in~\Cref{tab:watermark categories}.

\subsection{Pre-text Watermarks}

Pre-text watermarks can be categorized into token sampling-based and logits modification-based watermarks where the former modifies the token sampling strategy and the latter modifies logits during the inference phase.

\mypara{Token sampling-based watermark}
Watermark injection during token sampling is influenced by the randomness factor of the token sampling strategy.
The watermark injection phase utilizes a random number generated by a fixed seed or a random seed calculated based on prefix tokens to guide the token sampling procedure and generate watermarked texts.
During the watermark detection, alignment between output text tokens and the random number suffices for comparison.
In this paper, we explore three token sampling strategies: bit-string conversion~\cite{cryptoeprint:2023/763}, exponential~\cite{aaronson2022watermarking}, and inverse token sampling~\cite{kuditipudi2023robust}.

\begin{table}[t]
\centering
\caption{LLM watermarking schemes. Note that all the schemes here are black-box.}
\renewcommand{\arraystretch}{1.25} 
\begin{tabular}{p{2.4cm}p{2.2cm}p{2.6cm}}
\toprule 
\textbf{Category} & \textbf{Injection Method}  & \textbf{Scheme Name} \\ \midrule
\multirow{5}{2.3cm}{Pre-text Watermark} &\multirow{3}{2.5cm}{Token Sampling} & Convert~\cite{cryptoeprint:2023/763} \\
& &Inverse~\cite{kuditipudi2023robust}  \\
& &Exponential~\cite{aaronson2022watermarking} \\ \cmidrule{2-3}
& \multirow{2}{2.5cm}{Logits Modification} & KGW~\cite{kirchenbauer2023watermark} \\ 
& & Unigram~\cite{zhao2023provable} \\ 
\midrule
\multirow{3}{2.5cm}{Post-text Watermark} &\multirow{2}{1.7cm}{Format} & WHITEMARK~\cite{sato2023embarrassingly} \\ 
& & UniSpaCh~\cite{UniSpaCh} \\ \cmidrule{2-3}
&Lexical & Linguistic~\cite{yang2023watermarking} \\
\bottomrule
\end{tabular}
\label{tab:watermark categories}
\end{table}

\begin{itemize}
\item \textbf{Convert bit-string (Convert)}~\cite{cryptoeprint:2023/763} aims to develop undetectable watermarks.
This watermark can be viewed as the hash operation on the generated text.
Specifically, they first encode the generated text into a string with binary bits.
Then, they use a \{0,1\} hash function $h$ and sample tokens $x_j$ with preference for those satisfying $h(x_j) = 1$.
More tokens should hash to 1 than to 0 in watermarked text.
The probability of a token being encoded to 1 is defined as $p_j(1)$.
The watermarked model will output $x_j = 1$ if $u_j \leq p_j(1)$ and $x_j = 0$ otherwise.
Here $u_j$ is a real number sampled from [0,1].
The probability that the watermarked model output $x_j=1$ is $p_j(1)$.
For detection, they will compute a score $s(x_j,u_j)=\ln\frac{1}{u_j}$ if $x_j=1$ and $s(x_j,u_j)=\ln\frac{1}{1-u_j}$ otherwise.
Given a string $(x_1,...,x_L)$, the sum scores of detection is defined as $score=\sum_{j=1}^L s(x_j,u_j)$.
The score of the watermarked text should be significantly higher than the score of the normal text.
\item \textbf{Exponential sampling (Exponential)}~\cite{aaronson2022watermarking} injects the watermark to text by selecting a token $x_n$ that maximizes the score which depends on the probability $Pr(x_n|x_{0:n-1})$ and a pseudo-random value.
The pseudo-random value is defined as:
\begin{equation}
r_i = f(x_{n-H:n-1}, i),
\label{r-i}
\end{equation}
where $r_i \in [0,1]$ and $H$ presents the length of prior tokens.
To detect watermarks, a p-value which is define as $\sum_{j=1}^L \ln \frac{1}{1-r_{j'}}$ is also needed.
Here $r_{j'}$ is calculated by the token of output text on \Cref{r-i}, and $L$ is the token sequence length of output text.
We finally test whether it exceeds the threshold in the setting to implement the detection.
\item \textbf{Inverse token sampling (Inverse)}~\cite{kuditipudi2023robust} is similar to exponential. This strategy calculates the watermark scores from the next position of the watermark key $\xi$, $\xi = \xi^{(1)},\xi^{(2)},...,\xi^{(m)}$ where $\xi^{(i)} \in [0,1]^{\lvert \mathcal{V} \rvert}$ and m should exceed the max length of generated text, $\mathcal{V}$ is the vocabulary of tokens.
Finally, the watermark strategy is defined as:
\begin{equation}
    f(p,x,\xi)= onehot(\mathop{arg~max}\limits_{i}(\xi_i^{len(x)})^{1/p_i})
\end{equation}
where $x$ is the generated token string and $p$ is the probability distribution of the next token.
At the detection process, we perform a test statistic which is defined as:
\begin{equation}
    \phi(x,\xi)=\sum_{t=1}^{len(x)} (-\log(1-\xi_{x_t}^{(t)})
\end{equation}
where $t$ represents the token position in $x$.
This watermark is supposed to enhance the robustness compared to the Exponential.
It also computes a minimum Levenshtein distance using the test statistic as a cost and compares it to the expected distribution under the null.
\end{itemize}

\mypara{Logits modification-based watermark}
The prior watermarking schemes primarily address the challenge of achieving effective watermark injection and detection without adversely impacting the distribution of the model's output.
Building upon the aforementioned thesis, most watermarking schemes concentrate solely on modifying the watermark sampling strategy, often overlooking the influence of logits.
Kirchenbauer et al.~\cite{kirchenbauer2023watermark} propose a novel approach by introducing a bias to the logits, thereby altering the distribution of model output while maintaining output quality.
In this paper, we delve into two methods falling within this category which are proposed by Kirchenbauer et al.~\cite{kirchenbauer2023watermark} and Zhao et al.~\cite{zhao2023provable}.

\begin{itemize}
\item  \textbf{KGW.}
Kirchenbauer et al.~\cite{kirchenbauer2023watermark} propose a watermarking scheme without retraining the model. Specifically, during the inference process, a secret key can be calculated by the previous token and we can use the secret key to partition the vocabulary into a green list and a red list. Then, we add a bias to each green list logit before the Softmax function to calculate the probability distribution. We can detect the watermark by making a statistical test to the proportion of tokens in the text that belong to the green list.
Finally, the statistical measure of the z-score is defined as:
\begin{equation}
    z= \frac{|s|_G - \gamma L}{\sqrt{L(\gamma)(1-\gamma)}},
\end{equation}
where $|s|_G$ presents the number of green tokens, $L$ presents the total number of tokens in the text and $\gamma$ is a configuration parameter that presents the proportion of green list tokens in the whole vocabulary.
We consider the text to be watermarked if its z-score exceeds the pre-defined threshold.
\item \textbf{Unigram.}
Zhao et al.~\cite{zhao2023provable} introduce several enhancements to enhance the robustness of watermarks against removal attacks. Notably, they employ a fixed 1 prefix token for secret key computation, resulting in a consistent green list rather than a random one.
\end{itemize}

\subsection{Post-text Watermarks}

Post-text watermarks constitute a category of schemes wherein watermarks are appended to existing texts.
The predominant approach for this type of watermark involves the modification of the existing text to incorporate the watermarking information.
We implemented three watermark schemes belonging to this category in this paper, and we divided the three watermarks into two categories which are Format-based watermark and Lexical-based watermark according to the granularity of modifications.

\mypara{Format-based watermark}  
The Format-based watermark derives from an image watermark technology~\cite{article_image}.
It does not modify the content of text but changes the format of text that are imperceptible for humans, thereby implementing the watermark injection.

\begin{itemize}
\item \textbf{WHITEMARK}~\cite{sato2023embarrassingly} exploits the fact that the Unicode of text has several codepoints for whitespace. They can replace the origin whitespace (e.g. U+0020) with another new whitespace (e.g. U+2004) without degrading the quality of the text. The watermark can be detected by calculating the probability of new whitespace. 
\item \textbf{UniSpaCh}~\cite{UniSpaCh} propose a text-based data hiding method for Microsoft Word documents.
In their method, they leverage the whitespace between words and replace it with different Unicode space characters without degrading text quality. This approach can be considered a format-based watermarking technique for text.
During the detection process, they assess the number of replaced space characters and compare the results against a predefined threshold.
\end{itemize}

\mypara{Lexical-based watermark} 
This category of watermarks applies word-level modifications by replacing the random sampling words with other candidate synonyms without affecting the structure of the whole sentence.
This category of watermarks will be more imperceptible for humans but will induce a large consumption of time and computing resources.

\begin{itemize}
\item \textbf{Linguistic}~\cite{yang2023watermarking} first convert the token word into the binary bit based on an encoding function and the encoding bit-string conforms to the Bernoulli distribution where the probability of bit 1 is 0.5. Then, they selectively replace the words that represent bit 0 with the synonymous words that represent bit 1 to inject the watermark. The watermark can be detected by a statistical test.
\item \textbf{Generation-based watermark} Previous methods of watermark based on existing texts are based on the rule of text~\cite{sato2023embarrassingly,yang2023watermarking,munyer2024deeptextmark,article} which requires complex human design. Zhang et al.~\cite{zhang2023remark} propose a novel watermark module without human design which consists of three components: (\romannumeral1) Message encoding module; 
(\romannumeral2) Reparameterization module; (\romannumeral3) Message decoding module. The existing text $T = \{w_0,w_1,...,w_{n-1}\}$ and a binary watermark signature $s$ is denoted as input. The message encoding module outputs a watermarked distribution which is $P(T+s)$. $P(T+s)$ will go through the Reparameterization module and the Message decoding module which generates the watermark $s^*$. Watermark detection is achieved by comparing $s$ and $s^*$. 
However, our preliminary trial on~\cite{zhang2023remark} shows poor watermarking performance with different hyperparameter settings.
Therefore, we do not consider this method in our evaluation.
\end{itemize}

\begin{table}[t]
\centering
\caption{Watermark removal attacks.}
\renewcommand{\arraystretch}{1.25} 
\begin{tabular}{p{2.6cm}p{3.0cm}p{1.6cm}}
\toprule 
\textbf{Category} & \textbf{Attack Method} & \textbf{Access} \\
\midrule
\multirow{2}{2.6cm}{Pre-text Attacks} &Emoji attack~\cite{kirchenbauer2023watermark,goodside2023adversarial}&Black-box\\
&Distill~\cite{gu2024learnability} & White-box\\
\midrule
\multirow{10}{2.6cm}{Post-text Attacks} &Contraction~\cite{liang2022holistic} & \multirow{10}{1.6cm}{Black-box}\\
& Expansion~\cite{liang2022holistic}&\\
&Lowercase~\cite{liang2022holistic}& \\
&Misspelling~\cite{liang2022holistic}&\\
&Typo~\cite{liang2022holistic}&\\
&Modify~\cite{piet2023mark} & \\
&Synonym~\cite{piet2023mark}& \\
&Paraphrase~\cite{krishna2023paraphrasing} & \\
&Translation~\cite{finlay2021argos} & \\
&Token~\cite{kuditipudi2023robust} & \\
\bottomrule
\end{tabular}
\label{tab:attack categories}
\end{table}

\begin{table*}[t]
    \caption{The correspondence between watermark removal attacks and LLM watermarking schemes. Based on the respective mechanisms of the attacks and schemes, ``\ding{52}'' indicates that the attack can be applied against the corresponding watermarking scheme, while ``\ding{56}'' indicates that it cannot.}
    \centering
      \renewcommand{\arraystretch}{1.25} 
    \begin{tabular*}{
    \textwidth}{@{\extracolsep{\fill}}>{\centering\arraybackslash}p{1.9cm}>{\centering\arraybackslash}p{0.5cm}>{\centering\arraybackslash}p{0.9cm}>{\centering\arraybackslash}p{0.75cm}>{\centering\arraybackslash}p{1.1cm}>{\centering\arraybackslash}p{0.65cm}>{\centering\arraybackslash}p{1.4cm}>{\centering\arraybackslash}p{1.0cm}>{\centering\arraybackslash}p{1.4cm}}
    \toprule
    \multirow{2}{*}{\textbf{Attacks}} & \multicolumn{8}{c}{\textbf{Watermarking Schemes}} \\ \cmidrule{2-9}
    & KGW & Unigram & Inverse & Exponential & Convert & WHITEMARK & Linguistic & UniSpaCh\\
    \midrule
    Contraction &\ding{52} &\ding{52} &\ding{52}  &\ding{52}  &\ding{52}   &\ding{52}  &\ding{52}&\ding{52}   \\
    Lowercase &\ding{52} &\ding{52} &\ding{52} &\ding{52} &\ding{52} &\ding{52} &\ding{52} &\ding{52} \\
    Expansion &\ding{52} &\ding{52} &\ding{52} &\ding{52} &\ding{52}  &\ding{52} &\ding{52} &\ding{52} \\
    Misspelling &\ding{52} &\ding{52} &\ding{52} &\ding{52} &\ding{52}  &\ding{52} &\ding{52} &\ding{52} \\
    Typo &\ding{52} &\ding{52} &\ding{52} &\ding{52} &\ding{52}  &\ding{52} &\ding{52}&\ding{52}  \\
    Modify &\ding{52} &\ding{52} &\ding{52} &\ding{52} &\ding{52}  &\ding{52} &\ding{52} &\ding{52} \\
    Synonym &\ding{52} &\ding{52} &\ding{52} &\ding{52} &\ding{52}  &\ding{52} &\ding{52} &\ding{52} \\
    Token &\ding{52} &\ding{52} &\ding{52} &\ding{52} &\ding{52}  &\ding{52} &\ding{52} &\ding{52} \\
    Paraphrase &\ding{52} &\ding{52} &\ding{52} &\ding{52} &\ding{52}  &\ding{52} &\ding{52} &\ding{52} \\
    Translation &\ding{52} &\ding{52} &\ding{52} &\ding{52} &\ding{52}  &\ding{52} &\ding{52} &\ding{52} \\
    Emoji &\ding{52} &\ding{52} &\ding{52} &\ding{52} &\ding{56}  &\ding{56} &\ding{56} &\ding{56} \\
    Distill &\ding{52} &\ding{56} &\ding{52} &\ding{52} &\ding{56} &\ding{56} &\ding{56} &\ding{56} \\
    \bottomrule
    \end{tabular*}
    \label{tab: attack and watermark}
\end{table*}

\section{Watermark Removal Attacks}

We discuss twelve watermark removal attacks in this paper, in which two attacks belong to pre-text attacks, and the other ten attacks are post-text attacks (see Table~\ref{tab:attack categories}).

\subsection{Pre-text Attacks}

The Pre-text Attacks focus on the text generation process by introducing perturbations(e.g. emojis) prior to the generation of the text this category of attacks can only be applied to the Pre-text watermarks.
In addition, the distill attack will distill the target model without any effects on the generated text.
Hence, we also view this attack as a Pre-text Attack in our work.

\begin{itemize}
\item \textbf{Emoji attacks.} 
Kirchenbauer et al.~\cite{kirchenbauer2023watermark} propose an attack that aims to indiscriminate the generative capability of LLMs.
We can prompt the LLM to modify its output to some specific form that we desire.
Goodside~\cite{goodside2023adversarial} proposes an attack called emoji attack, which can prompt the LLM to generate an emoji after every word and then remove these emojis.
This attack will randomize the red list for subsequent tokens and thus can remove the watermark.
\item \textbf{Distill attack.}
Model extraction attack can be defined as training a new substitute model with the data generated by querying the target model.
Gu et.al~\cite{gu2024learnability} also propose to train a student model to behave like its teacher model.
Specifically, for the watermarking schemes that modify the logits, given an input text $x$, they will train the student model's next token distribution to match the watermarked teacher model.
For the watermarking schemes that modify the token sampling strategy, they first collect the watermarked text dataset from the teacher model and then fine-tune the student model on the dataset using the standard language modeling cross-entropy loss.
We name this attack method as Distill.
\end{itemize}

\subsection{Post-text Attacks}

Piet et al.~\cite{piet2023mark} implement a benchmark that evaluates the watermark capabilities of large language models.
There are several perturbations to the prompt in their code which are proposed in helm~\cite{liang2022holistic} and we select the perturbations that have no effect on the meaning of the text.

\begin{itemize}
\item \textbf{Contraction attack.}
The attack contracts verbs in the text (e.g. contract \textit{is not} to \textit{isn't}). 
\item \textbf{Expansion attack.}
In contrast to the contraction attack, this attack will expand some verb~(e.g. expand \textit{don't} to \textit{do not}).
\item \textbf{Lowercase attack.}
This attack converts all letters to lowercase.
\item \textbf{Misspelling attack.}
This attack misspells certain words with the probability $p$.
\item \textbf{Typo attack.}
This attack converts certain words to typos with the probability $p$.
\end{itemize}

Meanwhile, we also discuss the following post-text removal attacks:

\begin{itemize}
\item \textbf{Modify attack.}
For each word in the text, we can duplicate, remove, or replace the word with another word with probability $p$. 
\item \textbf{Synonym attack.}
This attack aims to replace the word with another semantically equivalent word with probability $p$.
Specifically, we follow the method which is implemented by Piet et al.~\cite{piet2023mark}.
We use WordNET to zero-prompt Llama-2 but not GPT-3.5 to generate alternative synonyms.
\item \textbf{Paraphrase attack.}
This is an attack that exploits another (paraphrase) language model to rephrase the existing texts. 
Because of the requirement of accessing another language model, this attack can be expensive to deploy for custom adversaries.
Krishana et al.~\cite{krishna2023paraphrasing} propose a language model that is fine-tuned specially for implementing paraphrase called dipper. 
In addition, we can also exploit the other models which belong to the Llama family (e.g.,
WizardLM-2-7B~\cite{xu2023wizardlm} or Llama-3-8B-instruct~\cite{llama3modelcard} to implement this attack. 
\item \textbf{Translation attack.}
Besides the generative language model, we can use a translation model to translate the existing text to another language and then translate the text back to the origin language (e.g., English to Russian to English).
In this paper, we use the Argos-translate model~\cite{finlay2021argos} based on OpenNMT~\cite{klein-etal-2017-opennmt}.    
\item \textbf{Token attack.}
Kuditipudi et al.~\cite{kuditipudi2023robust} propose a novel attack that modifies the output texts in the token dimension. 
We name it ``Token'' attack.
In this attack, the output texts will be encoded to the token list again.
Following the specified distribution for the token list, we sample tokens from this distribution randomly and the number of sampled tokens is determined by probability $p$.
The sampled tokens are subject to operations such as replacement, deletion, or iterative insertion.
\end{itemize}

\begin{table*}[h]
\centering
\caption{Metrics we used during evaluation.}
\label{tab:metric_intro}
\renewcommand{\arraystretch}{1.25} 
\begin{tabular}{llp{11cm}}
\toprule
\textbf{Metric}  &  \textbf{Symbol} & \textbf{Usage}\\
\midrule
Quality Score ($\uparrow$)   &  $Q \in [0,1]$ & To measure the quality of the (original or post-attacking) LLM-generated texts, thereby analyzing the performance of the removal attacks. \\ 
\midrule
Watermark Rate ($\uparrow$)  & $W \in [0,1]$ & To evaluate the extent of watermark information extraction.\\ \midrule
Robustness Score ($\uparrow$)  &  $R \in [0,1]$ & A comprehensive metric to evaluate the robustness of watermark schemes under removal attacks through combining $Q$ and $W$.\\ 
\bottomrule
\end{tabular}
\label{tab:watermark_requirements}
\end{table*}

\section{Experiment Settings}

\subsection{Basic Setup}

\mypara{Target model}
We use Llama-2-7B-chat~\cite{touvron2023llama} as our target model to implement all watermarking schemes.
Note that during inference, we use the below chat template to feed prompts. 
The max output sequence length is set to 1,024 tokens.

\begin{tcolorbox}[colback=gray!25!white, size=title,boxsep=1mm,colframe=white, after={\vskip0mm}]
\tiny

\texttt{<s>[INST]<<SYS>>\{\{system\_prompt\}\}<</SYS>}

\textit{Question: \{question\}}\texttt{[/INST]} 

\textit{Answer:}

\end{tcolorbox}

\mypara{Watermark generation dataset}
\label{sec: quality}
All the watermarking schemes discussed in this paper leverage the same dataset, i.e., watermark generation dataset ($\mathcal{D}_{\rm Gen}$), to inject and detect watermark information. 
For instance, we use the dataset\footnote{\url{https://github.com/wagner-group/MarkMyWords/blob/main/src/watermark_benchmark/utils/generation_prompts.py}.} which proposed by ~\cite{piet2023mark} as our $\mathcal{D}_{\rm Gen}$.
$\mathcal{D}_{\rm Gen}$ contains 296 instructions, covering three long text generation tasks (book report, story generation, and fake news).

\subsection{Evaluation Metrics}
\label{sec:metric}

As~\Cref{tab:metric_intro} shows, there are a total of three metrics used in our paper, that is, Quality Score, Watermark Rate, and Robustness Score.

\mypara{Quality score}
We use the quality score proposed by~\cite{piet2023mark} to evaluate the impact of removal attacks on the quality of AI-generated text, i.e., to quantify the change in text quality before and after the attack.
Following~\cite{liu2023geval,chiang2023large,Hackl_2023,wang2023chatgpt,kocmi2023large,chen2023exploring,piet2023mark}, we leverage LLM to evaluate the quality of each AI-generated text.
For instance, given a \textit{question} $x_i$ from $\mathcal{D}_{\rm Gen}$, we first feed it into the target model to generate its \textit{response}, then we feed them together with an evaluating prompt into Llama-3-8B-instruct~\cite{llama3modelcard}, one of the most advancing open-source model, to get a quality grade $q_i \in [0,1]$.
If the \textit{responses} do not suffer attacks, we calculate the mean quality score of the whole dataset as $Q^{clean}=(\sum_i^{n} q_i)/n$, $n$ is $|\mathcal{D}_{\rm Gen}|$.
Similarly, we use $Q^{attack}$ to denote the mean quality score of the attacked responses.
As a result, the final quality score could be defined as:
\begin{equation}
    Q  = 0.5 \cdot Q^{clean} + 0.5 \cdot (\max(0,\min(\frac{Q^{attack}}{Q^{clean}},1))).
\end{equation}
Note that we use the prompt from~\cite{piet2023mark} as our evaluating prompt, i.e, 

\begin{tcolorbox}[colback=gray!25!white, size=title,boxsep=1mm,colframe=white, after={\vskip0mm}]
\tiny
\texttt{<|begin\_of\_text|><|start\_header\_id|>\textbf{system}<|end\_header\_id|>}

\textit{You are given a question and a response, and you provide a grade ranging from [0,1] measuring the quality of the response, in terms of accuracy, level of detail, and typographical, grammatical, and lexical correctness. Remove points as soon as one of the criteria is missed. }

\textit{\#\#\# Question: \{question\}}

\textit{\#\#\# Response: \{response\}}

\texttt{<|eot\_id|><|start\_header\_id|>\textbf{user}<|end\_header\_id|>}

\textit{Please provide a grade.}

\textit{\#\#\# Grade:} 

\texttt{<|eot\_id|><|start\_header\_id|>\textbf{assistant}<|end\_header\_id|>}

\end{tcolorbox}

\mypara{Watermark rate}
Given the watermark generation dataset $\mathcal{D}_{\rm Gen}$, we use the watermark rate $W$ to calculate the proportion of watermark information that can be successfully extracted from the model output, i.e.,

\begin{equation}
    W = \frac{\sum_{i}^{|\mathcal{D}_{\rm Gen}|}\mathbf{1}\{\mathsf{Scheme}_{\rm dec}(y_i,\kappa; \mathcal{M})=\textit{True}\}}{|\mathcal{D}_{\rm Gen}|},
\end{equation}
where $y_i$ is the AI-generated text generated from $x_i \in \mathcal{D}_{Gen}$, i.e., $y_i=\mathsf{Scheme}_{\rm inj}(x_i,\kappa; \mathcal{M})$, and $\mathbf{1}(\cdot)$ is the indicator function.

\mypara{Robustness score}
We combine the quality score and the watermark rate to propose a comprehensive metric, i.e., robustness score $R$.
We aim to leverage $R$ to intuitively quantify the robustness of a watermarking scheme.
Given a single removal attack $a_i$, we define the robustness score of the watermarking scheme against $a_i$ as
\begin{equation}
\label{eq:robust_id}
    R_{a_i}  =  0.5 \cdot Q_{a_i} + 0.5 \cdot W_{a_i},
\end{equation}
where $Q_{a_i}$ and $W_{a_i}$ are calculated from the responses that have been attacked by the attack $a_i$.

We further consider that a watermarking scheme can suffer a set of attacks, i.e., $\mathcal{A}=\{a_1,...,a_m\}$.
Naturally, we could define the robustness score under such scenario as
\begin{equation}
R_{\mathcal{A}} = (\sum_i^m R_{a_i})/m.
\end{equation}

\subsection{Experimental Setup}

This section provides all the hyper-parameters and other experiments set up for all watermark and attack methods.

\mypara{Watermark setup}
We summarize the setup of all eight watermark schemes here.
For the Logits modification-based watermarks, in the experiments of KGW, we use the setting that has the best performance, which is $(\gamma,\delta) =(0.25,2)$, where $\gamma$ is the proportion of green list tokens on the whole vocabulary, $\delta$ is the bias we added to the logits of every token during the inference process.
In Unigram, we use the same setting as KGW which also achieves the best performance in our experiments.

For the Token sampling-based watermark schemes, in Convert and Exponential, the number of hashed keys is set to 4.
In Inverse, the number of hashed keys is set to 4 and the number of shifts is set to 2.

For the Format-based watermark, in WHITEMARK, the replaced whitespace is $U+2004$ and the probability of replacement is set to 0.6.
In UniSpaCh watermark, the several specific whitespace we used are \{\ $U+2000$, $U+2001$, $U+2004$, $U+2006$, $U+2007$, $U+2008$, $U+2009$, $U+200A$ \}.
The probability of replacement is set to 0.6.

For the Lexical-based watermark, we use the default setting that the similarity threshold is set to 0.5 and the number of candidate synonyms is set to 8 for every word.

Note that the watermark detection threshold is set to 0.95 for all watermark schemes.

\mypara{Attack based on LLMs setup}
The attack methods in our paper have been divided into two categories which are attack based on LLMs and attack based on existing text.
For the former category, we implement two attack methods including Emoji attack~\cite{goodside2023adversarial} and distill attack~\cite{gu2024learnability}.
For the latter category, we implement nine attack methods including contraction, lowercase, expansion, misspelling, typo which are proposed by Liang et al.~\cite{liang2022holistic}, modify, synonym which are proposed by Piet et al.~\cite{piet2023mark}, paraphrase~\cite{krishna2023paraphrasing} and translation~\cite{finlay2021argos}.

The former five attack methods have different experiment setups and we will introduce them respectively.
We implement the latter nine attack methods with the same setup. These attack methods will be applied to the model outputs where the watermark has been injected.

\mypara{Emoji attack}
During the inference phase, we execute this assault. For the query dataset, we append an attack prompt to each query, specifically instructing, “Additionally, please add two \emojismile emojis after each output word.”
Subsequently, all \emojismile emojis present in the model’s output will be eliminated.
This attack is capable of targeting the watermark while minimizing any potential degradation to the textual quality.

\mypara{Distill attack}
In this attack, we need to train a student model on the dataset which is distilled from the target teacher model.
Both the teacher and student models are Llama-2-7B-chat.
Different from the typical distill methods, this attack will teach the student model to match the next token distribution outputted by the teacher model when using watermark based on modifying the logits~(e.g. KGW~\cite{kirchenbauer2023watermark}, Unigram~\cite{zhao2023provable}).
For the watermark based on modifying the token sampling strategy~(e.g. Inverse~\cite{kuditipudi2023robust}, Exponential~\cite{aaronson2022watermarking}, Convert~\cite{cryptoeprint:2023/763}), we first query the model to generate watermarked samples and then fine-tune the student model on these watermarked samples.
We sample 5,000 questions from OpenWebText~\cite{Gokaslan2019OpenWeb} to form our distill attacking dataset. 

\mypara{Quality evaluation}
In our quality evaluation of the model outputs, we use the query prompt introduced in \Cref{sec: quality}.
We will consider the system prompt, model input, and output in this process so that the length will be much longer.
We set the max sequence length to 1536 tokens to meet this requirement for quality evaluation.

\mypara{Watermark detection}
The specific verification ways are different for different watermarking schemes. 
However, all verification ways take the model output as input so that we can have the unified setup which is introduced in Section~\ref{iniwatermark}.

\mypara{Multi-attack setup}
In addition to the initial single attack on the watermarking schemes, we propose novel attack methods that apply multiple attacks to the same watermark.
We first apply every two attack methods to the same watermarking schemes and we also use the same attack twice. We then calculate the quality and watermark rate after every combined attack and select some prominent attacks to apply the multiple attacks.
In this work, we screened out the attack methods with the best attack effects.

\section{Evaluation Results}

In this section, we first evaluate the robustness of the watermarks against different individual attacks, along with performing extensive experiments to assess the quality and watermark rate against the combined attack (i.e., with two or more different attack strategies combined together), from which we identify the optimal attack combination.
This combined attack was then subjected to further discussion and experimental analysis.
After that, we evaluate the efficiency and imperceptibility of different watermarks.

\begin{figure*}[!t]
\centering
\begin{subfigure}{0.48\columnwidth}
\includegraphics[width = \columnwidth]{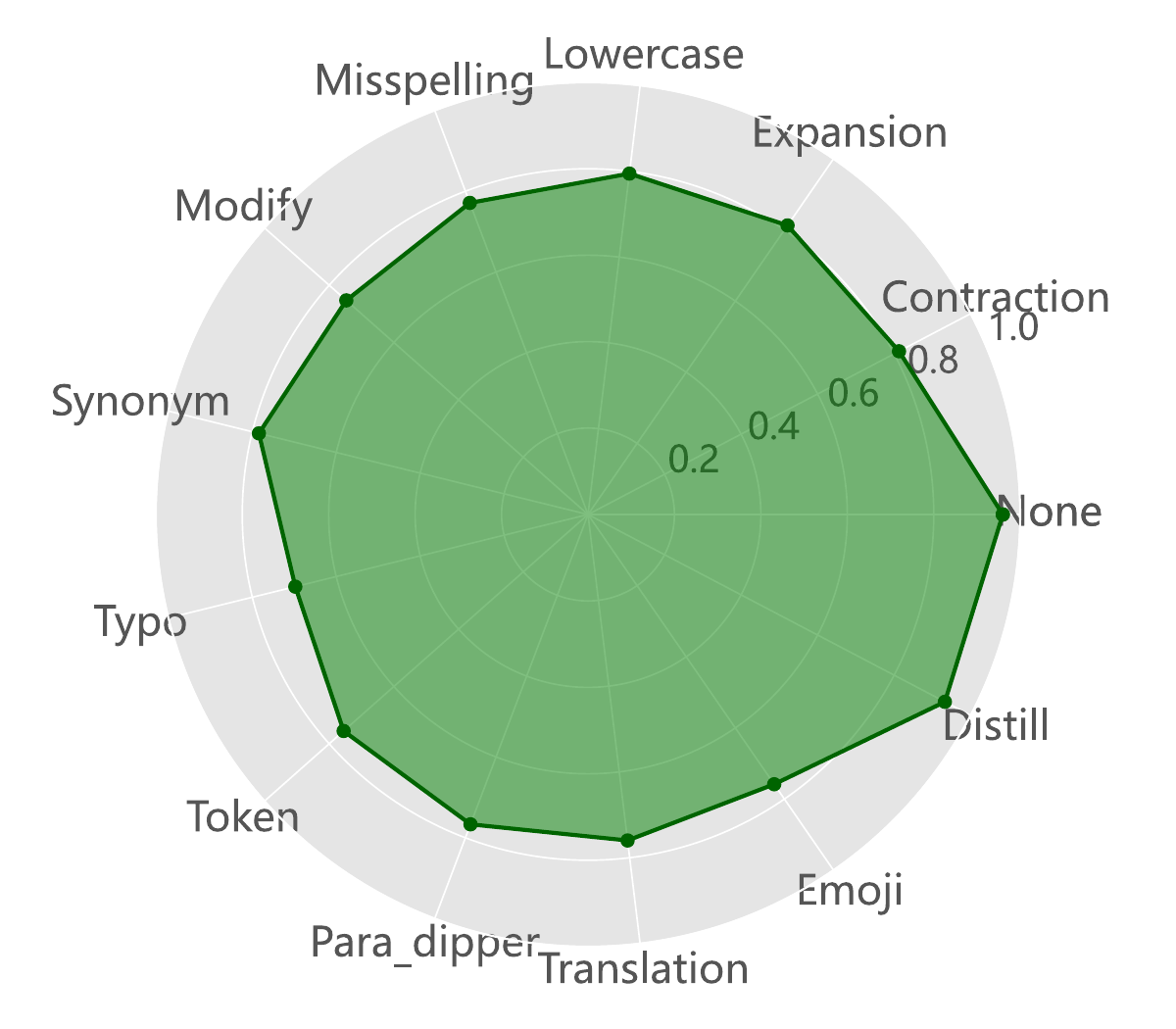}
\includegraphics[width = \columnwidth]{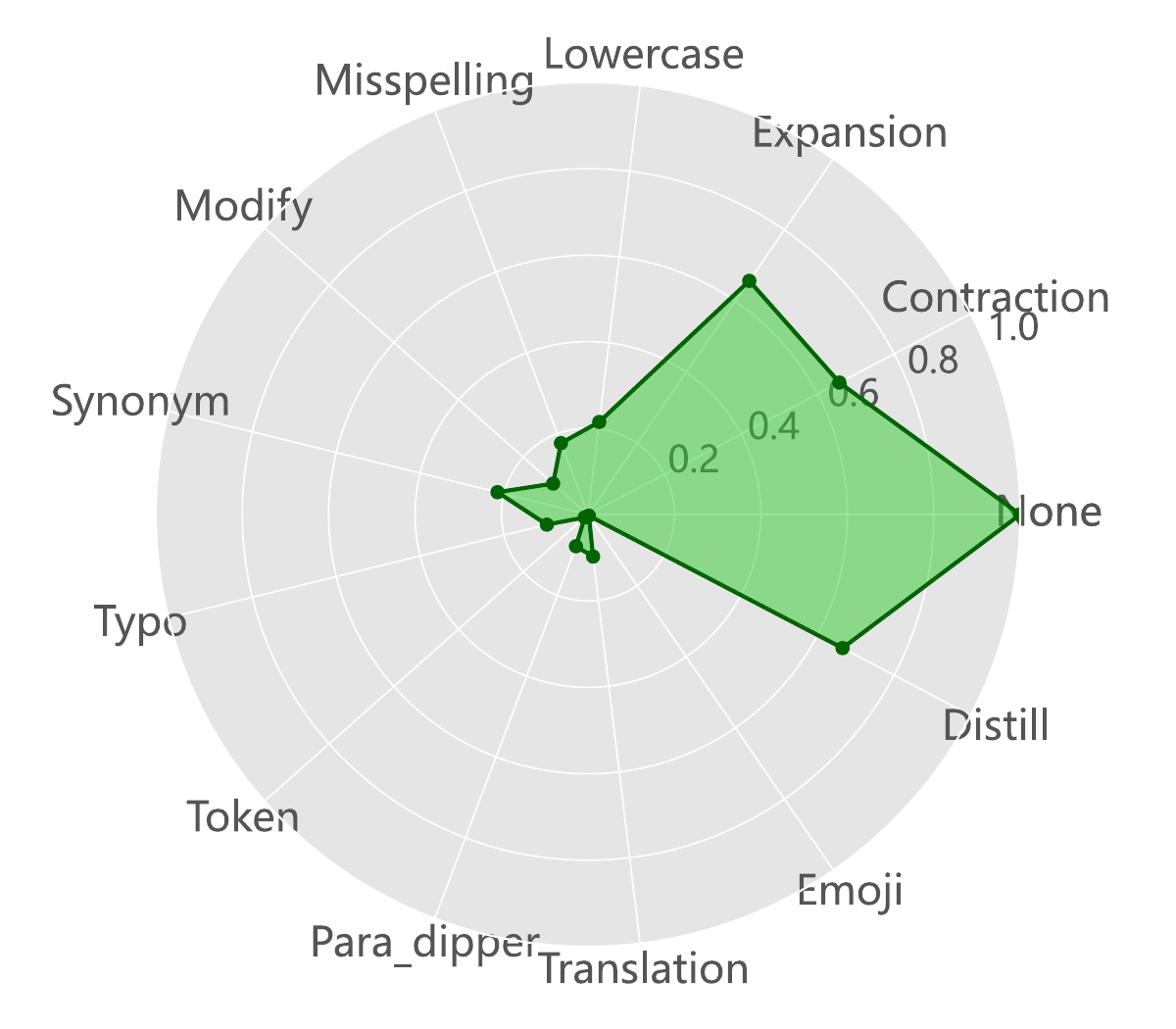}
\caption{KGW}
\label{KGW}
\end{subfigure}
\begin{subfigure}{0.48\columnwidth}
\includegraphics[width=\columnwidth]{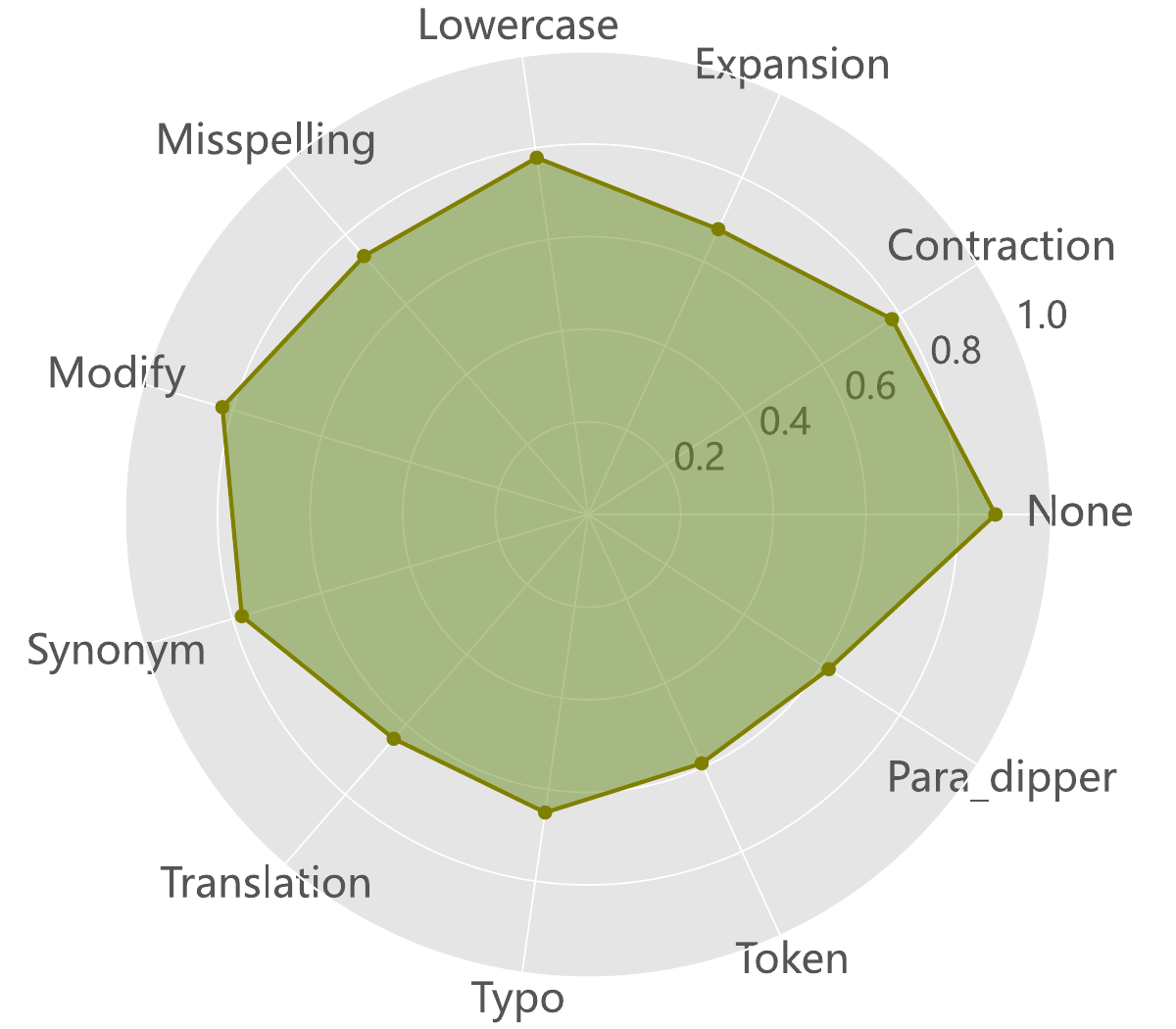}
\includegraphics[width=\columnwidth]{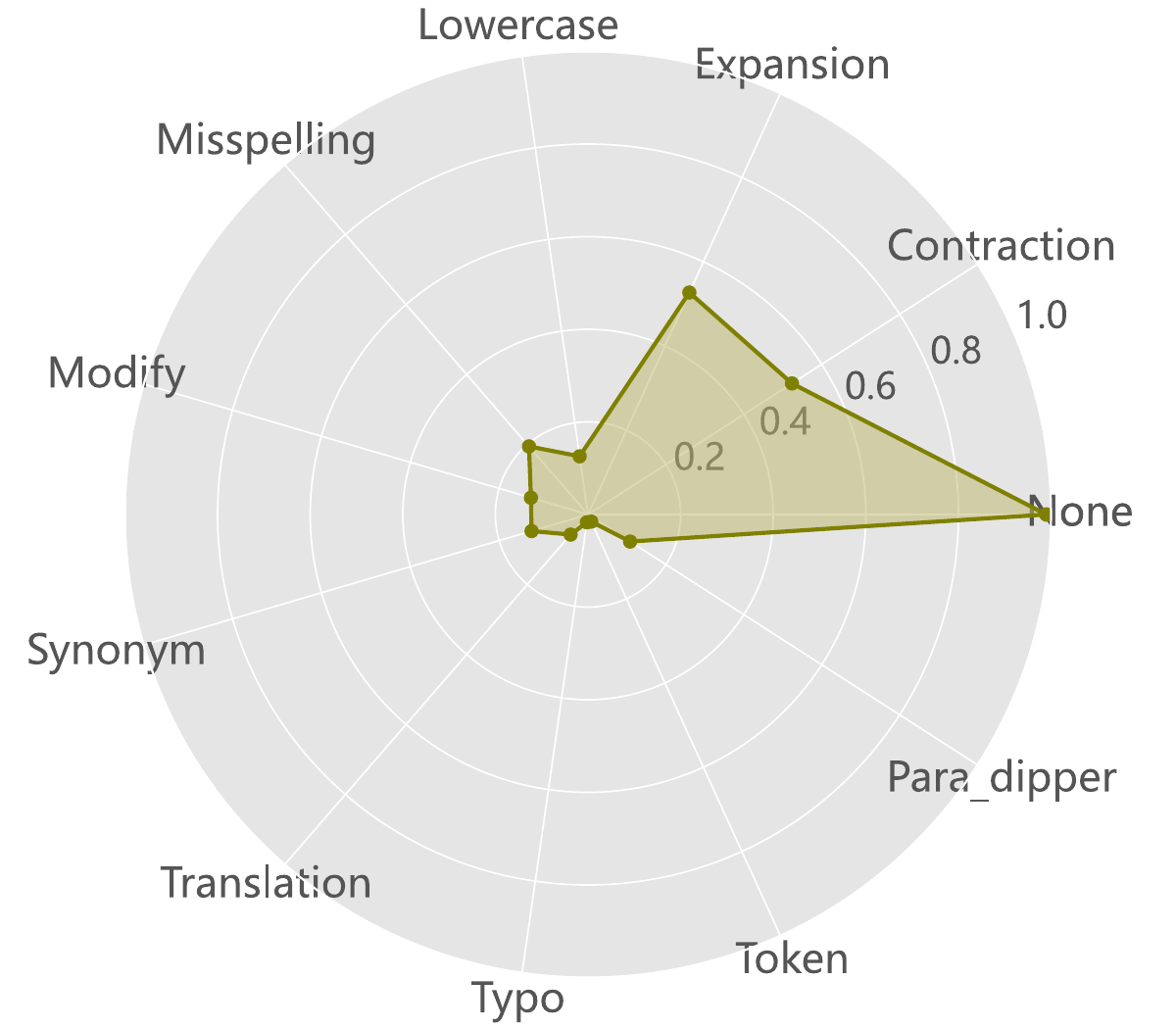}
\caption{Unigram}
\label{Unigram}
\end{subfigure}
\begin{subfigure}{0.48\columnwidth}
\includegraphics[width=\columnwidth]{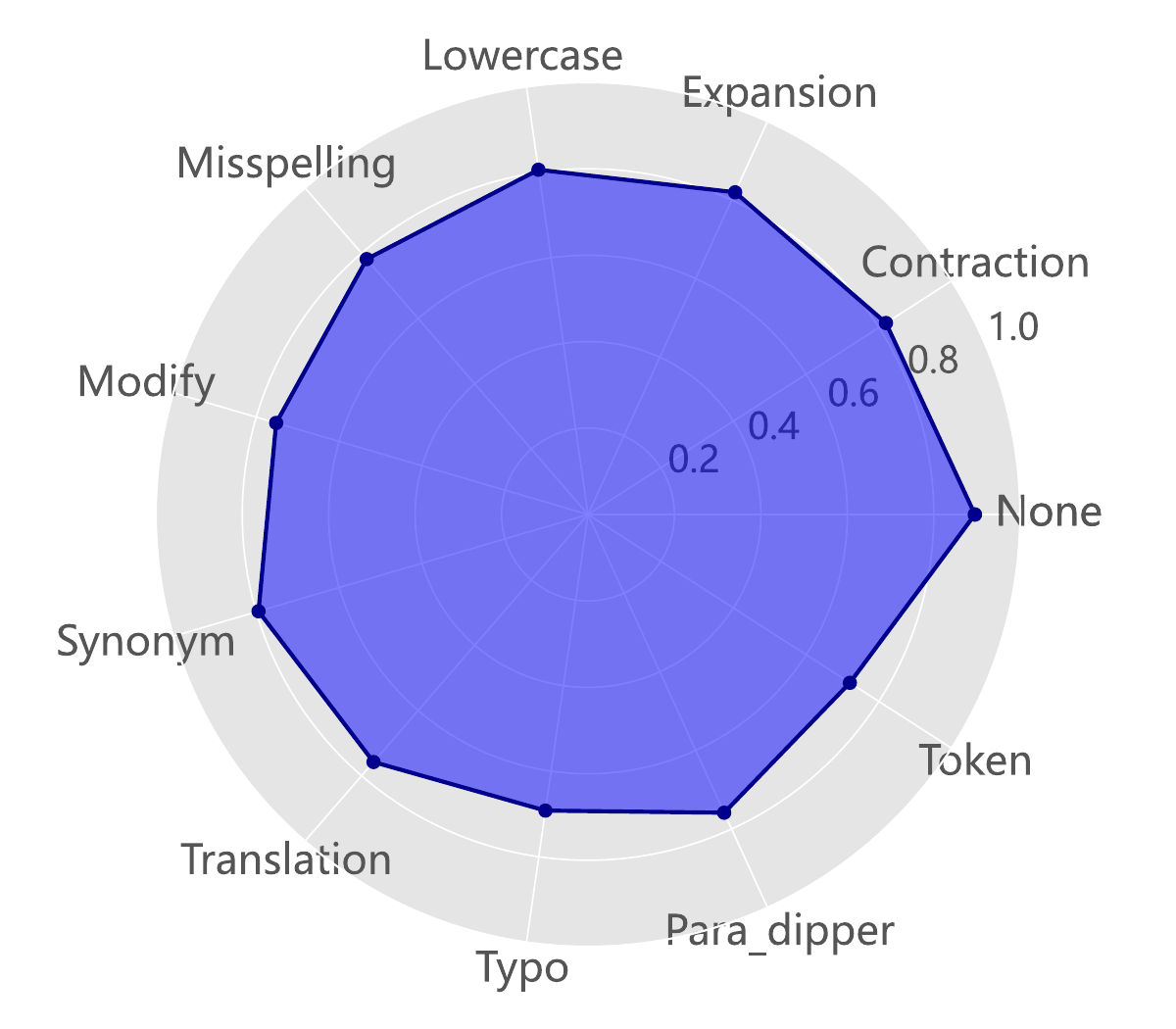}
\includegraphics[width=\columnwidth]{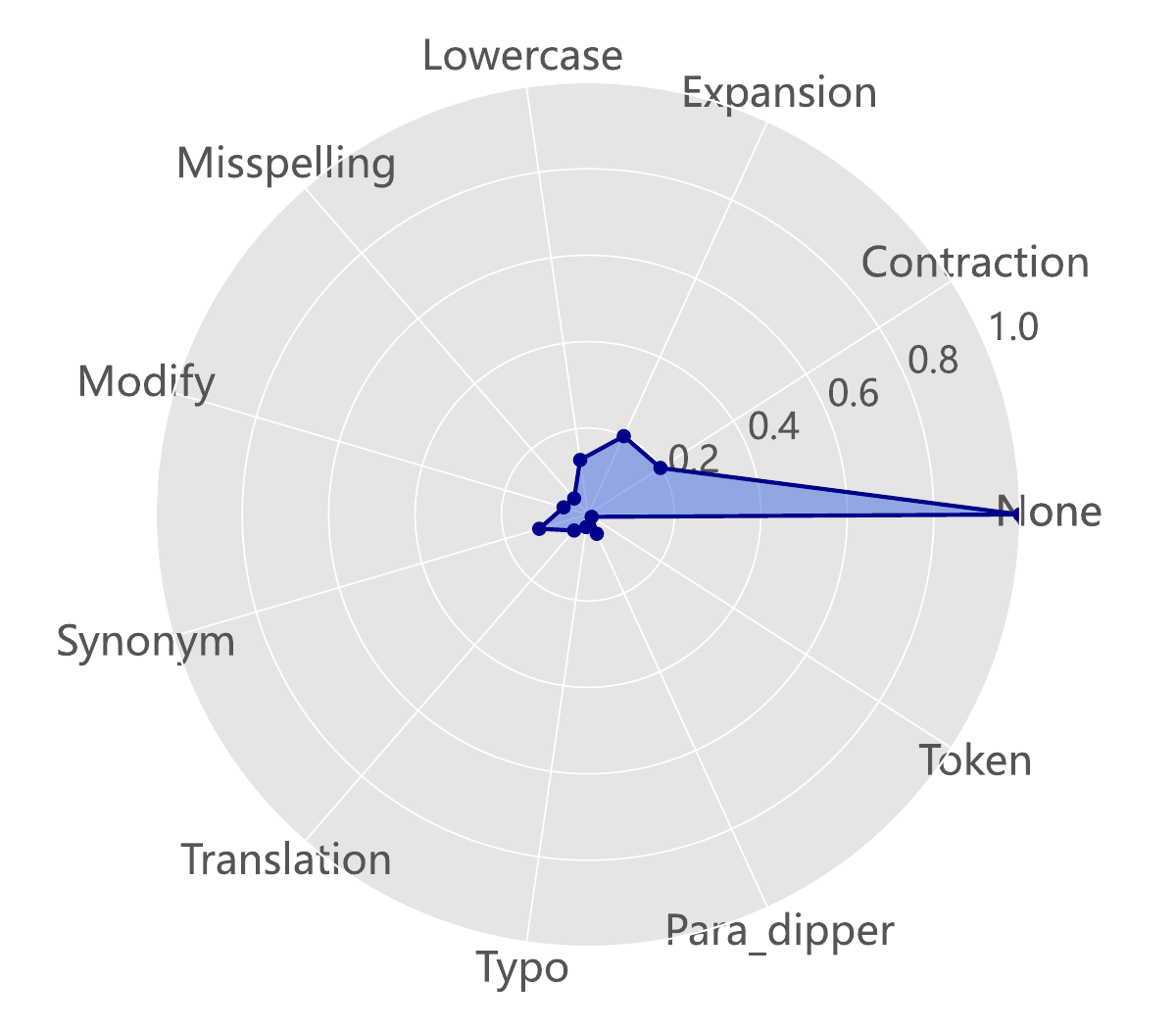}
\caption{Convert}
\label{Convert}
\end{subfigure}
\begin{subfigure}{0.48\columnwidth}
\includegraphics[width=\columnwidth]{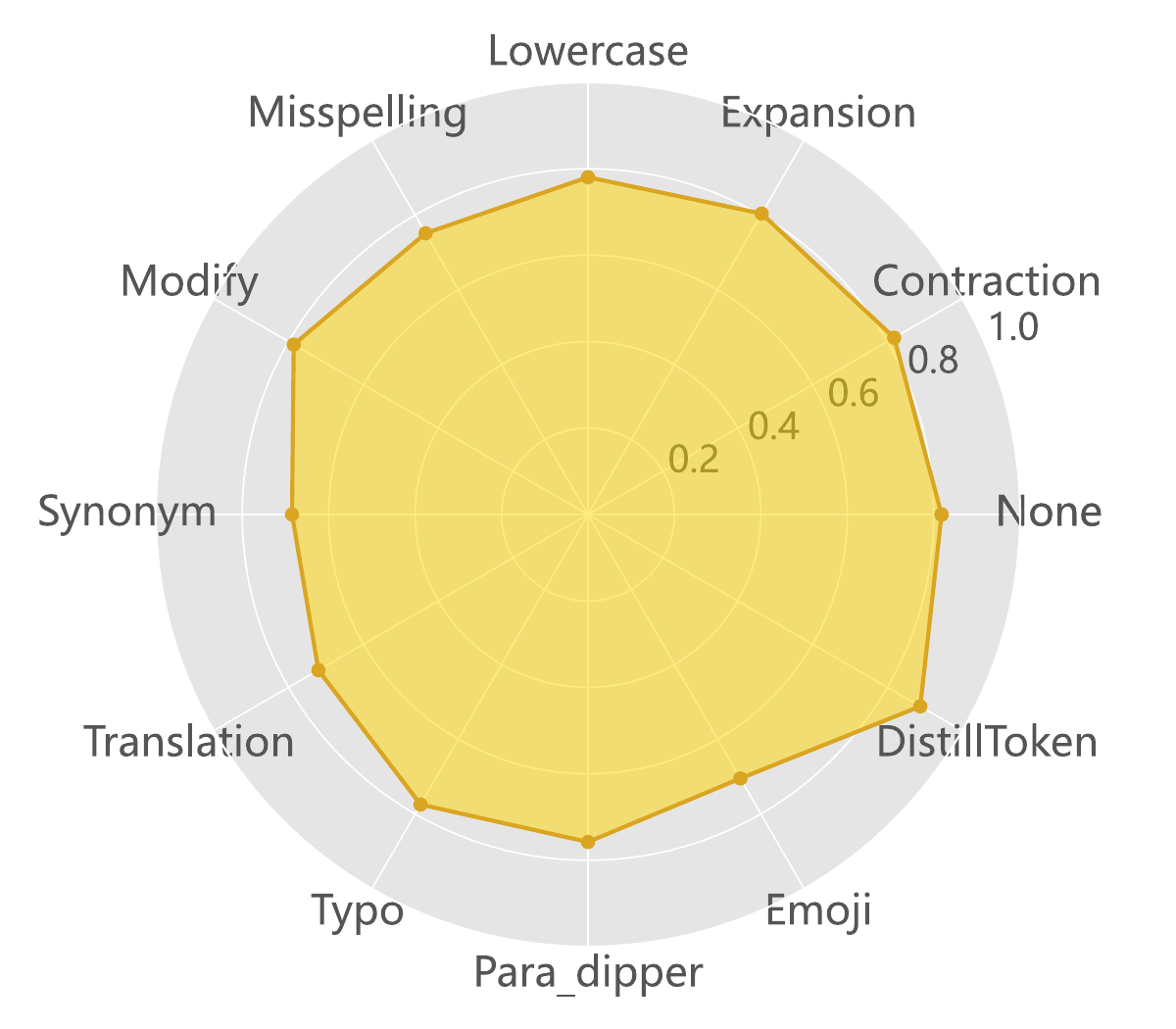}
\includegraphics[width=\columnwidth]{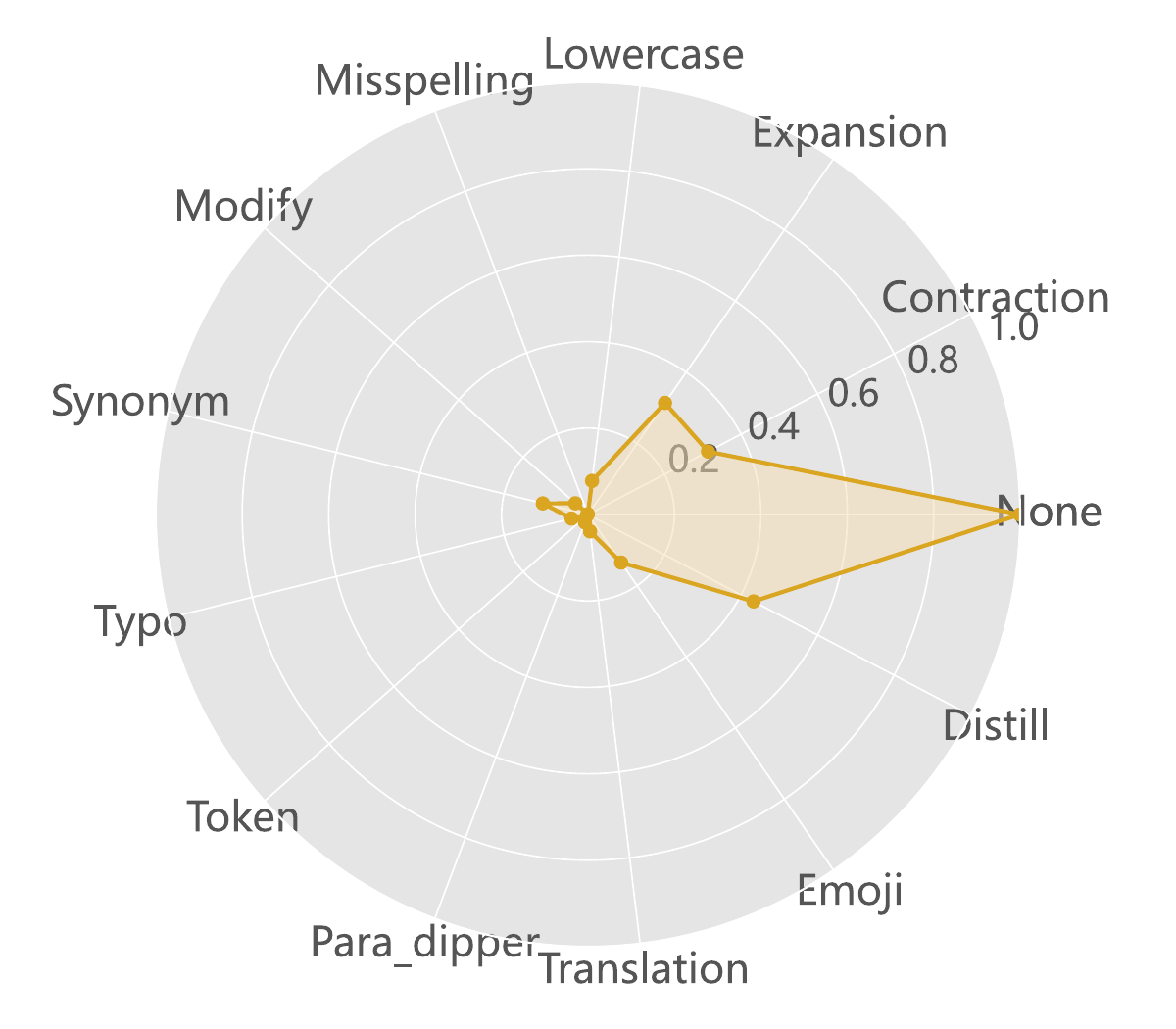}
\caption{Inverse}
\label{Inverse}
\end{subfigure}
\begin{subfigure}{0.48\columnwidth}
\includegraphics[width=\columnwidth]{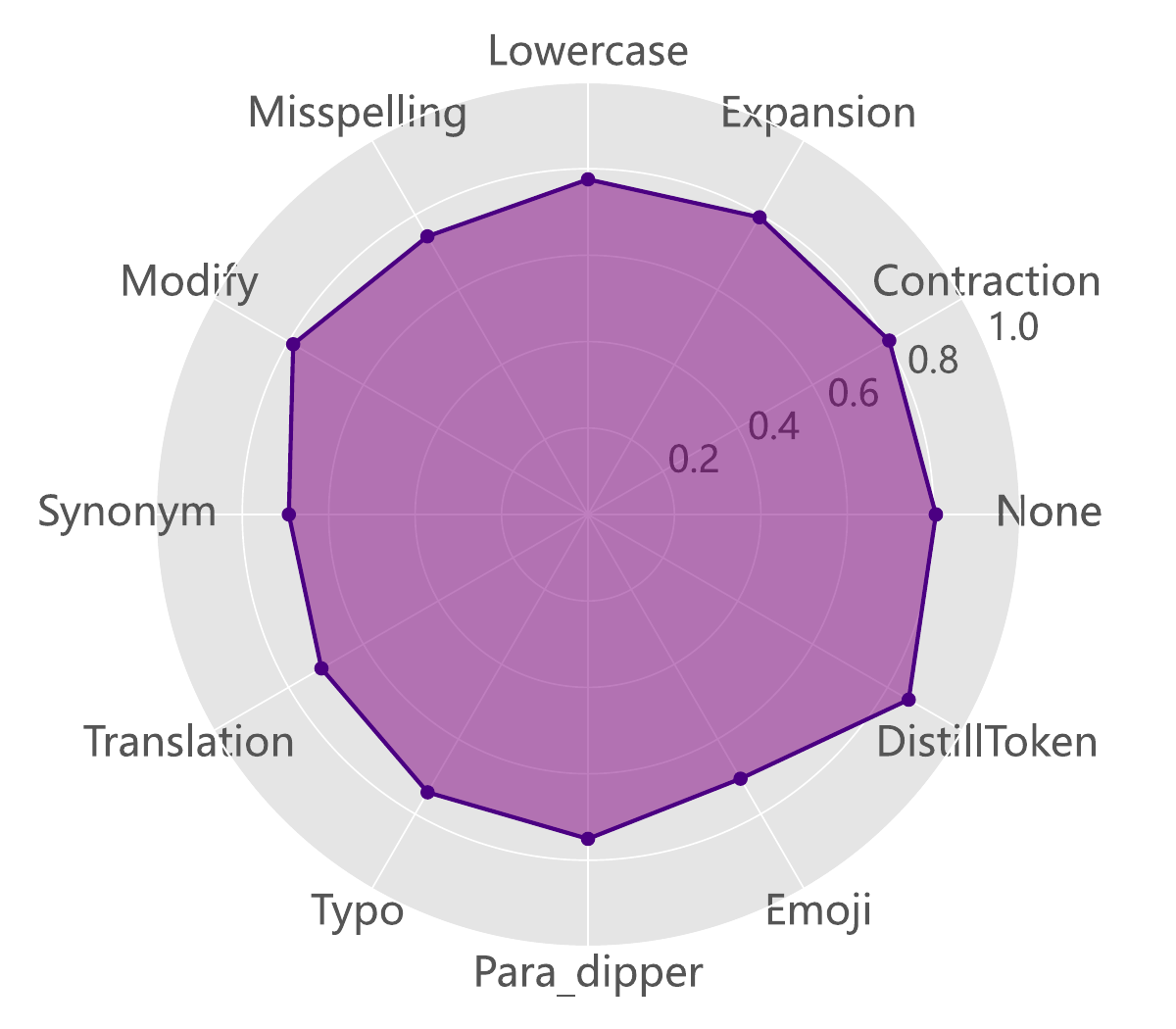}
\includegraphics[width=\columnwidth]{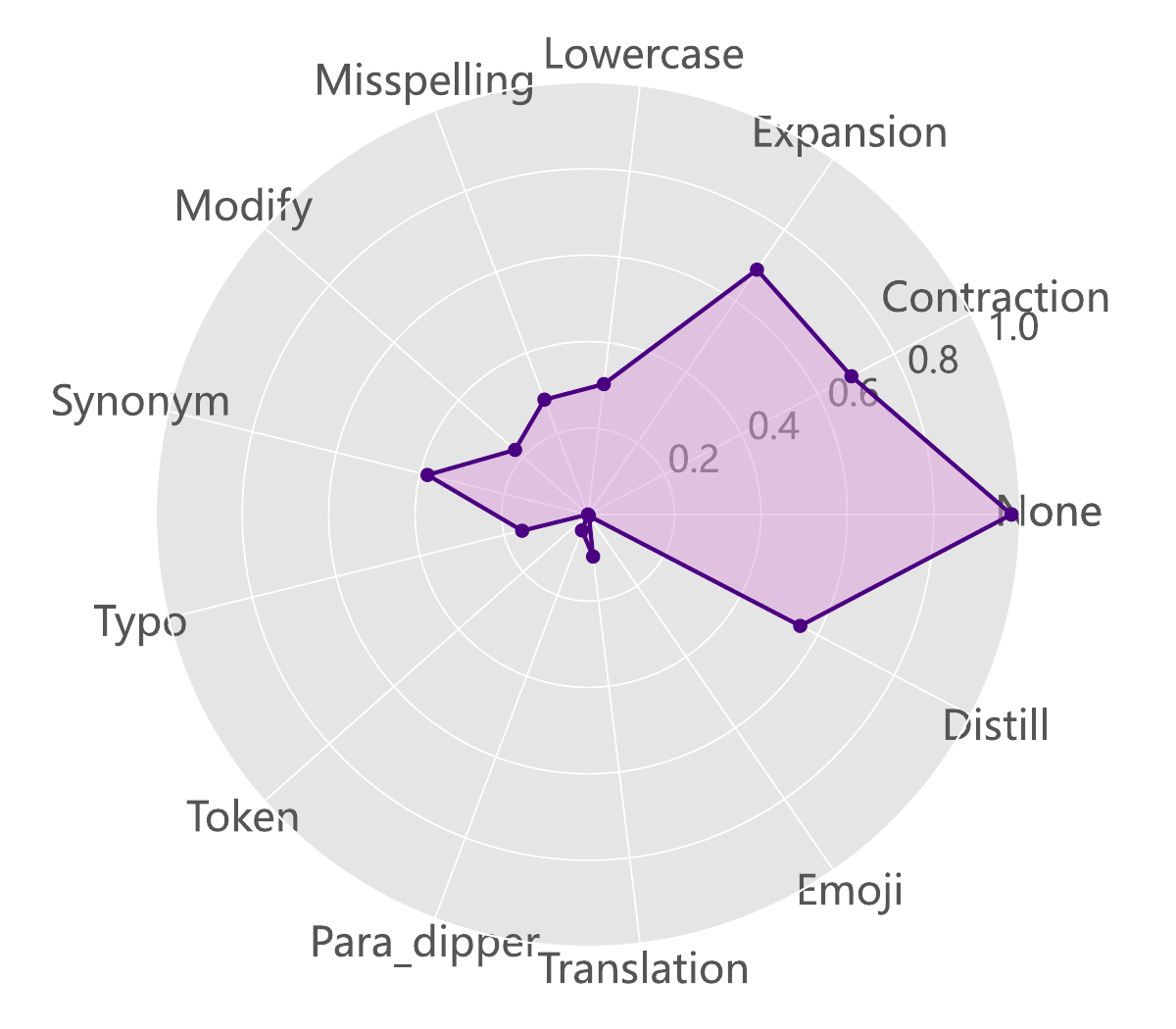}
\caption{Exponential}
\label{Exponential}
\end{subfigure}
\begin{subfigure}{0.48\columnwidth}
\includegraphics[width=\columnwidth]{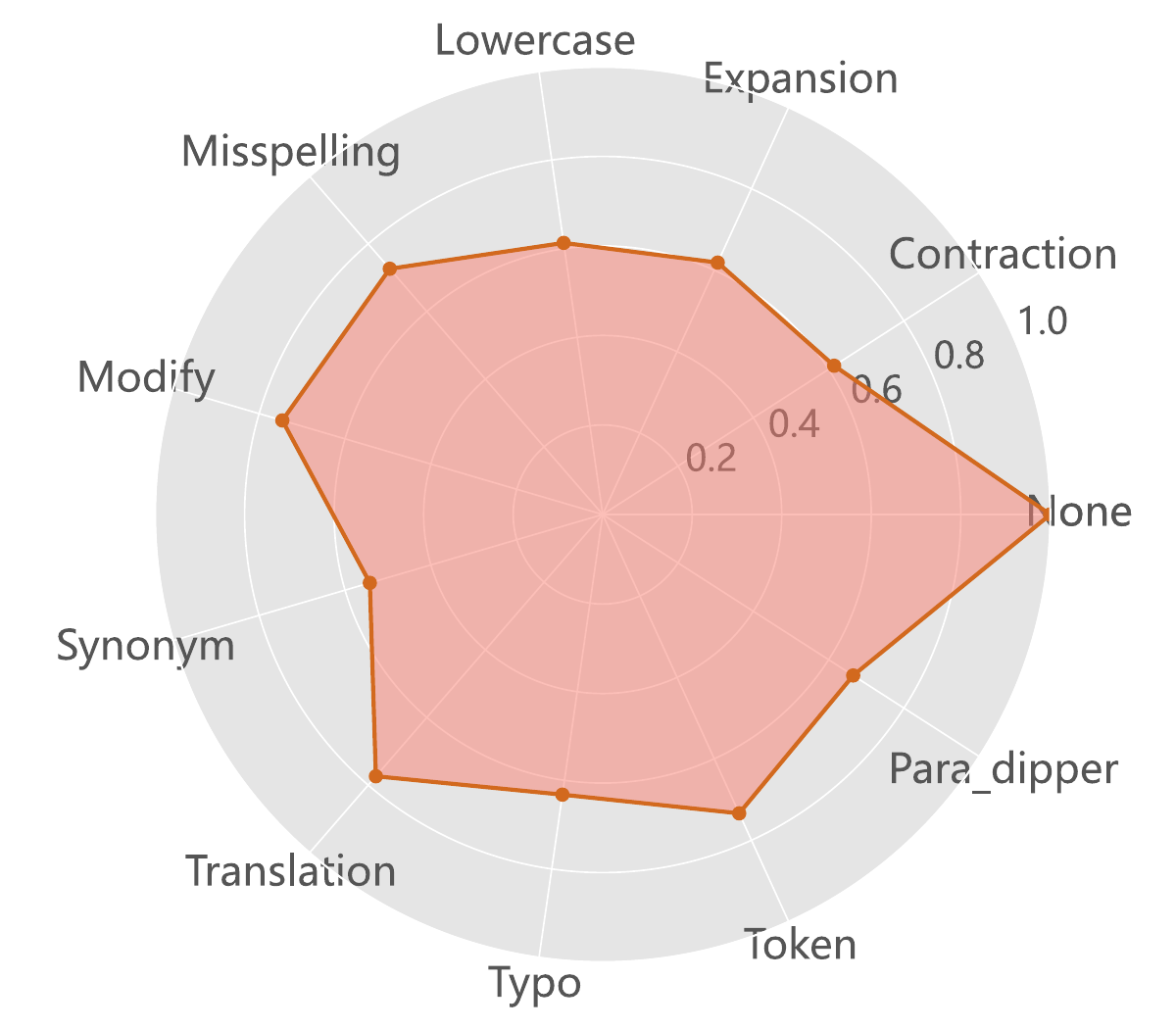}
\includegraphics[width=\columnwidth]{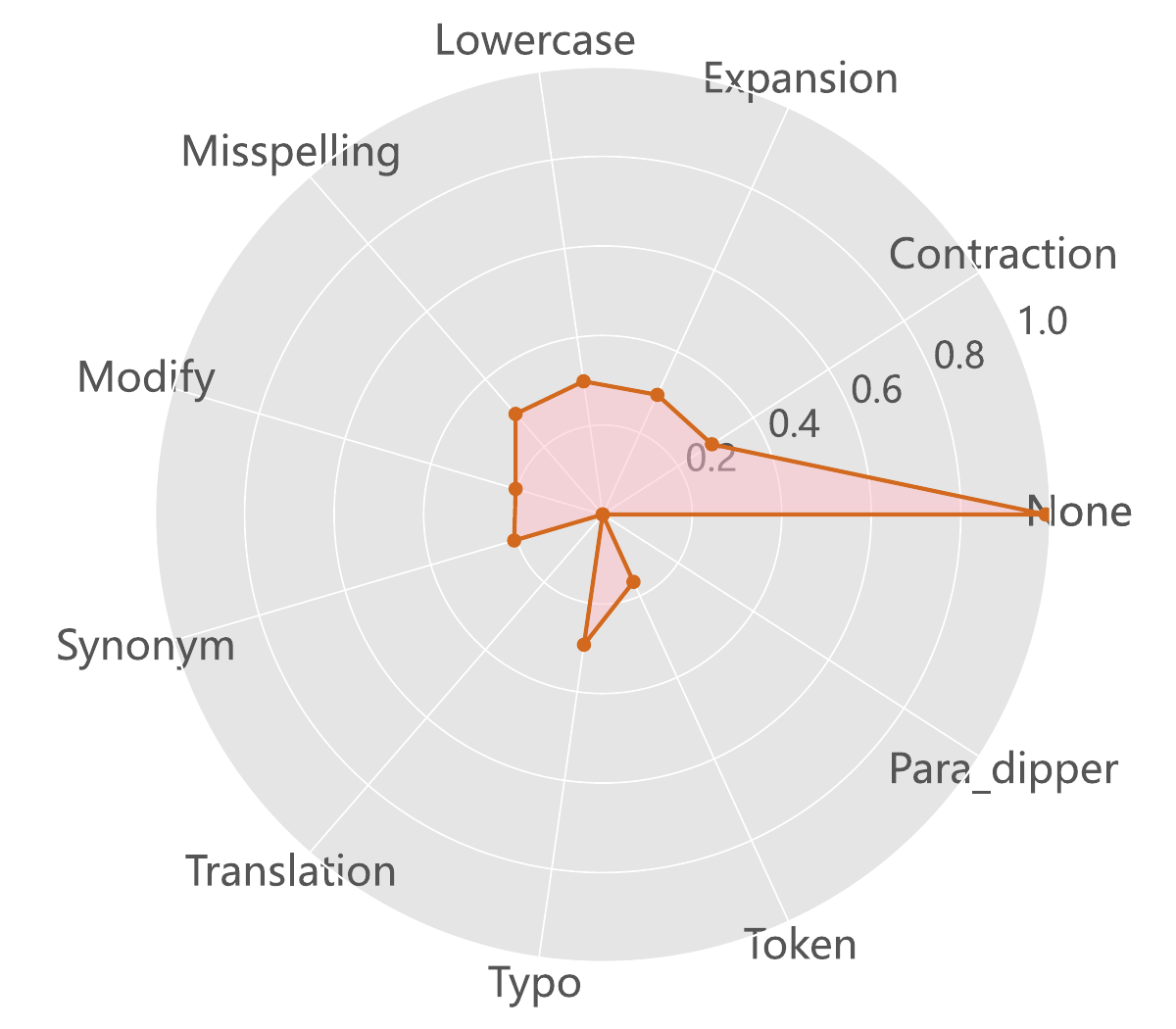}
\caption{WHITEMARK}
\label{White}
\end{subfigure}
\begin{subfigure}{0.48\columnwidth}
\includegraphics[width=\columnwidth]{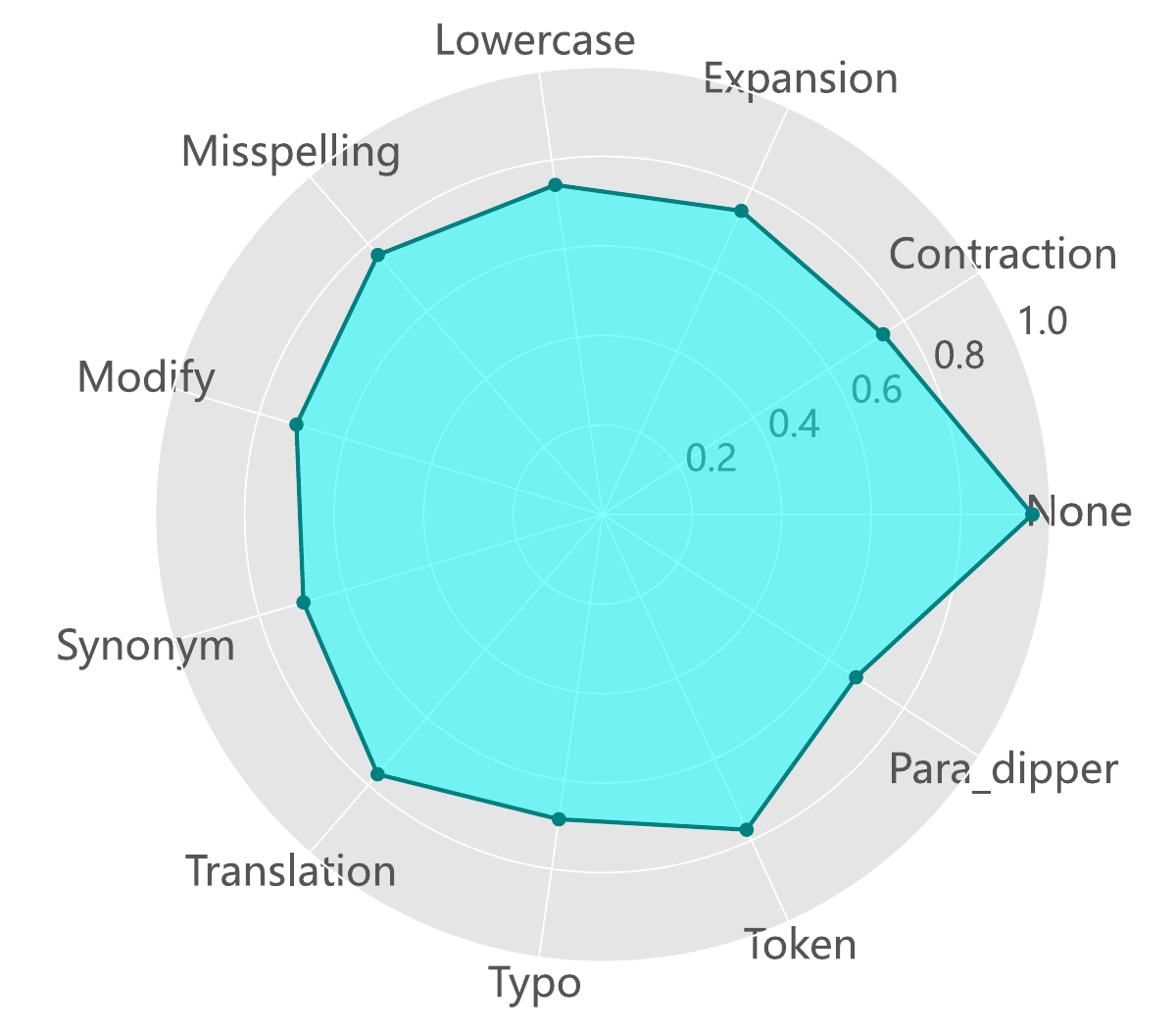}
\includegraphics[width=\columnwidth]{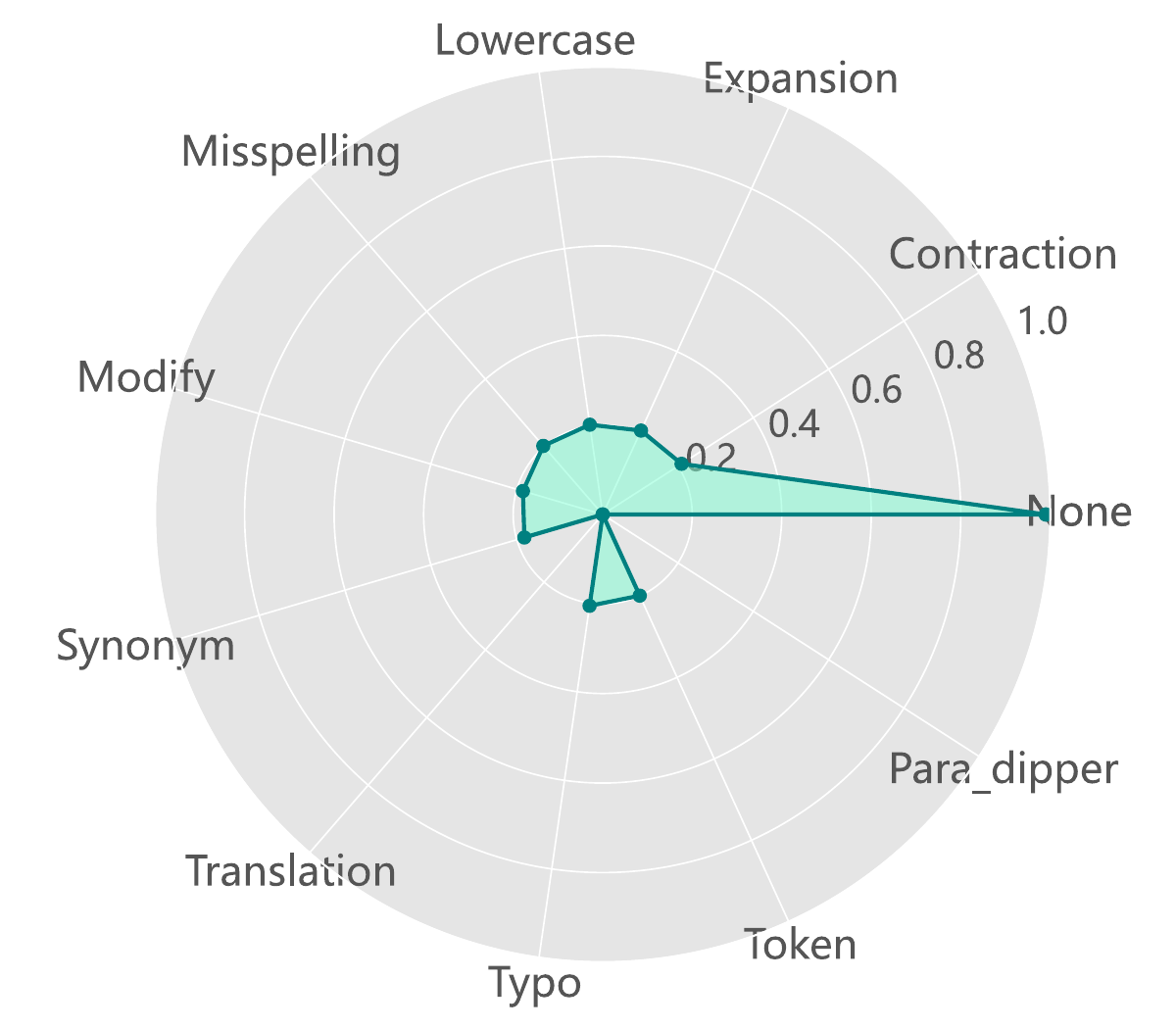}
\caption{UniSpaCh}
\label{UniSpaCh}
\end{subfigure}
\begin{subfigure}{0.48\columnwidth}
\includegraphics[width=\columnwidth]{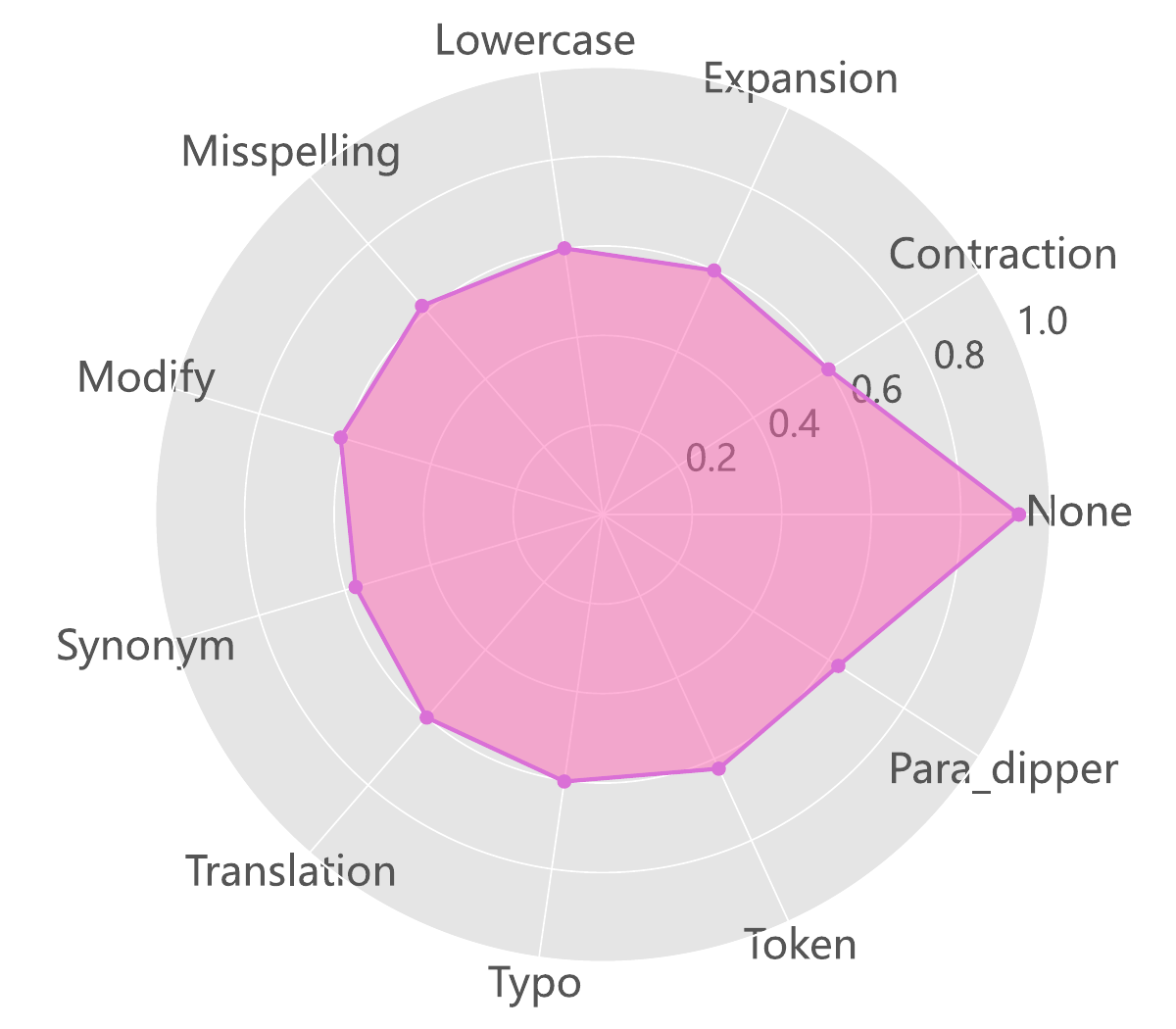}
\includegraphics[width=\columnwidth]{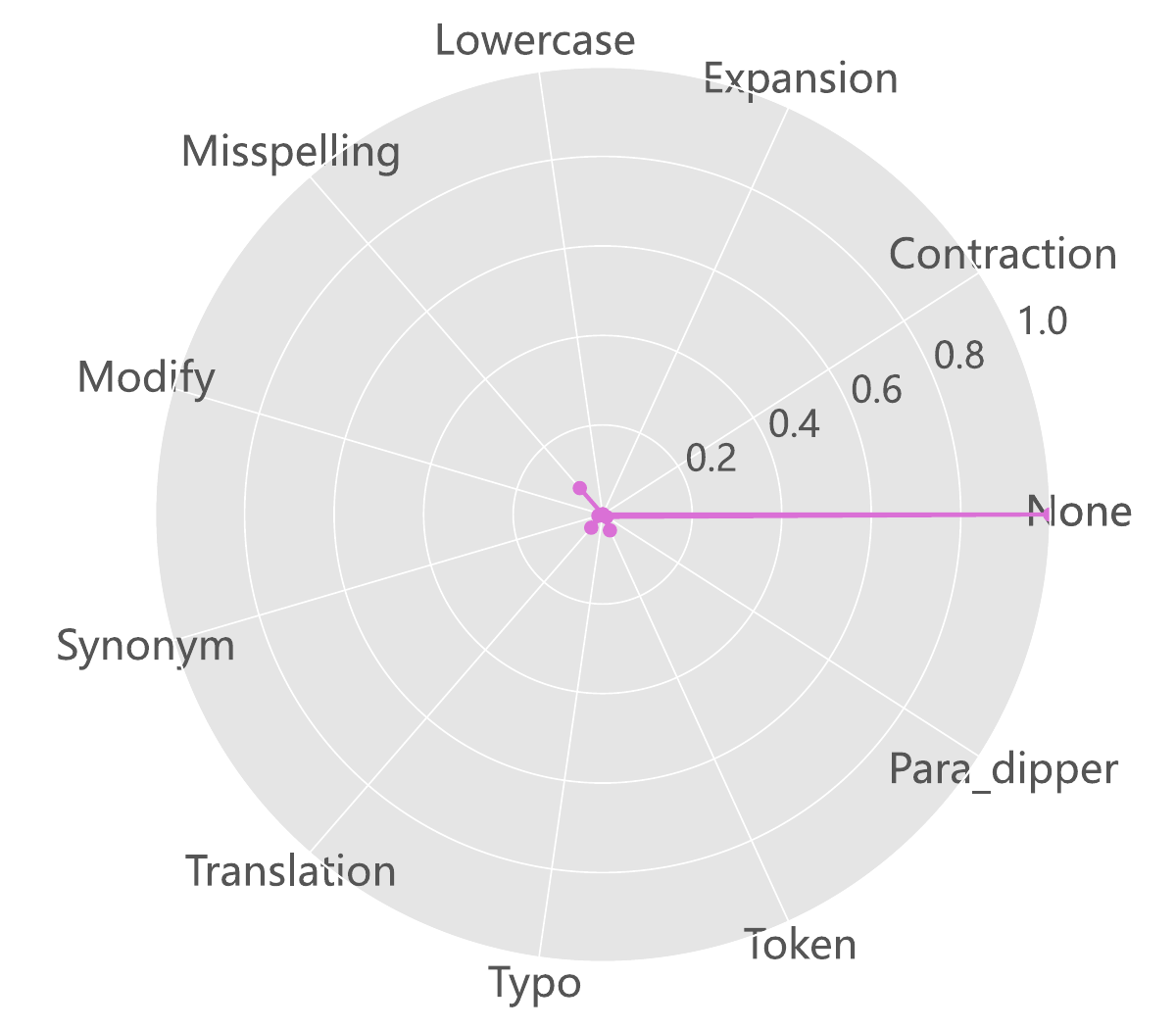}
\caption{Linguistic}
\label{Linguistic}
\end{subfigure}
\caption{The performance of quality (upper one of each subfigure) and watermark rate (lower one of each subfigure) for all watermarking schemes. \Cref{KGW}, \Cref{Unigram}, \Cref{Convert}, \Cref{Inverse}, and \Cref{Exponential} represent the results of pre-text watermarks. \Cref{White}, \Cref{UniSpaCh}, and \Cref{Linguistic} represent the results of post-text watermarks.}
\label{fig: results of watermark LLMs}
\end{figure*}

\begin{table}[h]
\centering
\caption{Robustness scores of different watermarking schemes against \textit{individual} removal attacks. Here we report the mean value of all post-text attacks or pre-text attacks. }
\begin{tabular}{p{2.0cm}p{1.2cm}p{1.2cm}p{1.0cm}}
\toprule
\multirow{2}{*}{\textbf{Watermarks}} & \multicolumn{3}{c}{\textbf{Removal Attacks}} \\ \cmidrule{2-4}
  & \textbf{Post-text Attacks}  & \textbf{Pre-text Attacks} & \textbf{Avg.} \\
\midrule
\rowcolor{gray!25}KGW    & 0.5095 & 0.5345 & \underline{\textbf{0.5220}}\\
Unigram  &0.4419 &0.4427 & \textbf{0.4423}\\
Convert & 0.4362&0.4166 & \textbf{0.4264}\\
Inverse  &0.4413 &0.4363 & \textbf{0.4454}\\
\rowcolor{gray!25}Exponential &0.5528 &0.4495 & \underline{\textbf{0.5012}}\\
\midrule
WHITEMARK  &0.4473 &0.3499 & \textbf{0.3986}\\
UniSpaCh  &0.4576 &0.3842 & \textbf{0.4209}\\
Linguistic & 0.3031&0.3317 & \underline{\textbf{0.3174}}\\
\bottomrule
\end{tabular}
\label{tab: robustness of all}
\end{table}

\subsection{Robustness Against Individual Attacks}

We first evaluate the robustness of different watermarking schemes against individual watermark removal attacks, including pre-text attacks and post-text attacks.

\mypara{Robustness score analysis}
We first launch individual attacks against watermarking schemes and report the mean values of the robustness scores (\Cref{eq:robust_id}).
As Table~\ref{tab: robustness of all} shows, we observe that pre-text watermarks are more robust than post-text watermarks in general.
For instance, KGW and Exponential reach $0.5220$ and $0.5012$ robustness scores, respectively, while WHITEMARK and UniSpaCh only reach $0.3986$ and $0.4209$ robustness scores.
This is expected as the pre-text watermarks usually involve more complex strategies to inject the watermark into the whole text while post-text watermarks only replace a few tokens, which makes them more vulnerable to attacks.
However, the robustness score falls significantly short of the benchmark established for the ideal watermark (which should be above 0.95 robustness score), indicating a need for further optimization and improvement.

\begin{figure*}[!t]
\centering
\begin{subfigure}{0.39\columnwidth}
\includegraphics[width=\columnwidth]{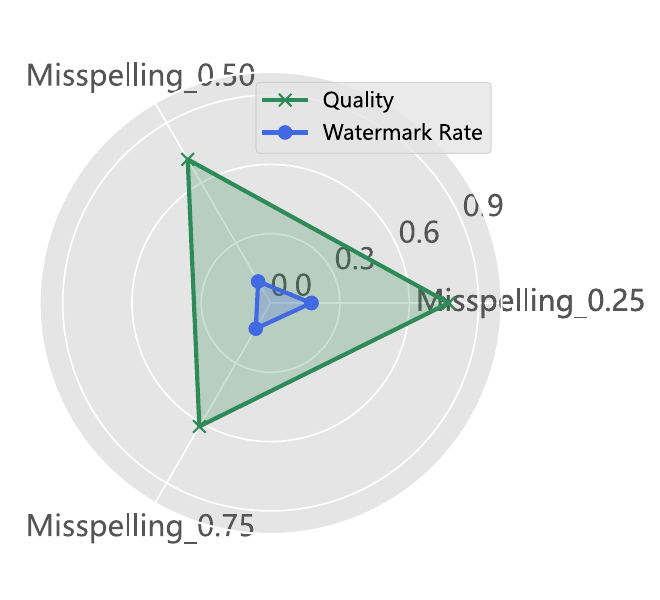}
\caption{Helm}
\label{Helm}
\end{subfigure}
\begin{subfigure}{0.39\columnwidth}
\includegraphics[width=\columnwidth]{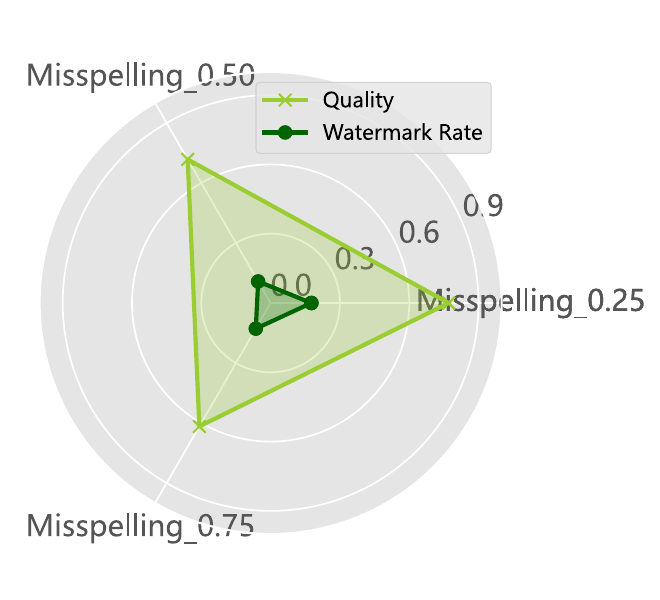}
\caption{Misspelling}
\label{Misspelling}
\end{subfigure}
\begin{subfigure}{0.39\columnwidth}
\includegraphics[width=\columnwidth]{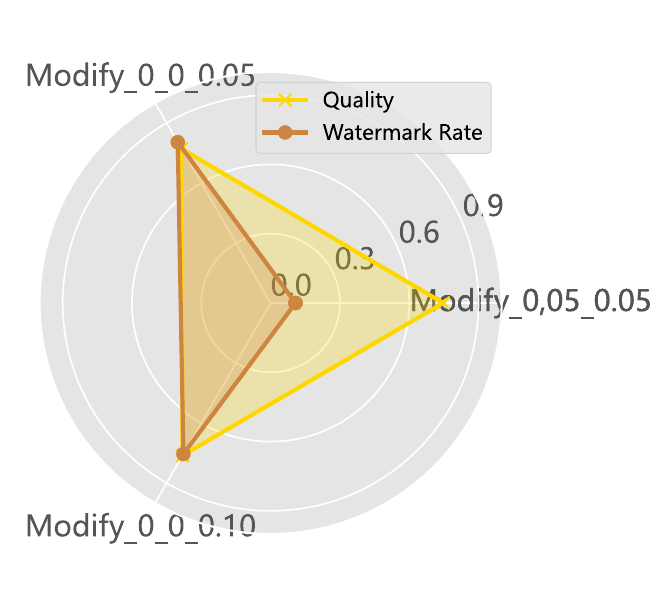}
\caption{Modify}
\label{fig:attack_modify_params}
\end{subfigure}
\begin{subfigure}{0.39\columnwidth}
\includegraphics[width=\columnwidth]{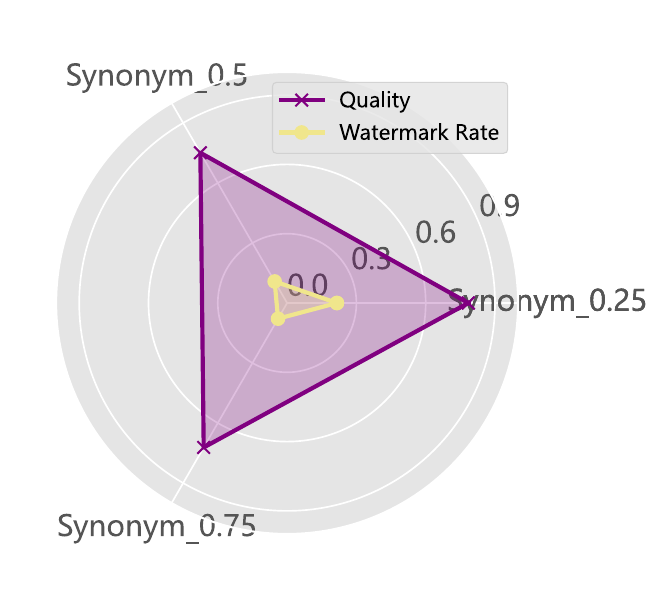}
\caption{Synonym}
\label{Synonym}
\end{subfigure}
\begin{subfigure}{0.39\columnwidth}
\includegraphics[width=\columnwidth]{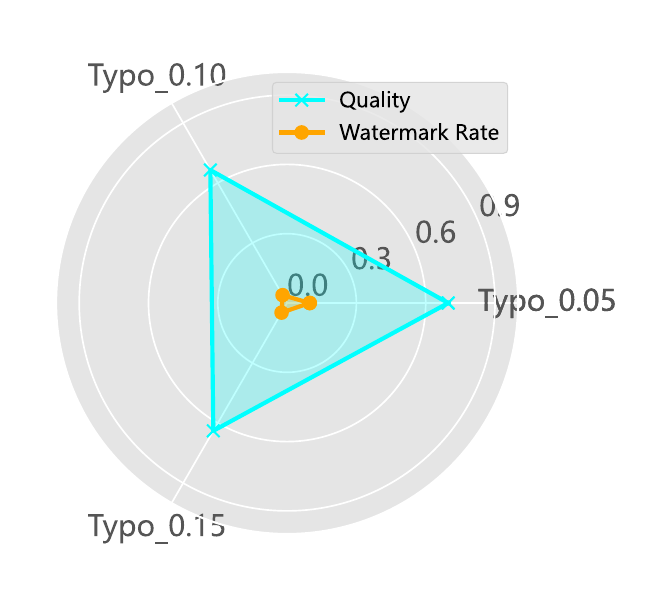}
\caption{Typo}
\label{Typo}
\end{subfigure}
\begin{subfigure}{0.39\columnwidth}
\includegraphics[width=\columnwidth]{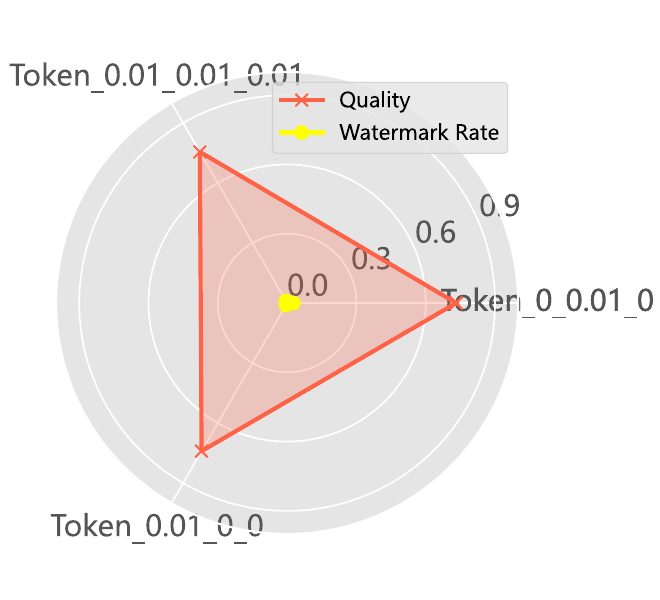}
\caption{Token}
\label{Token}
\end{subfigure}
\begin{subfigure}{0.39\columnwidth}
\includegraphics[width=\columnwidth]{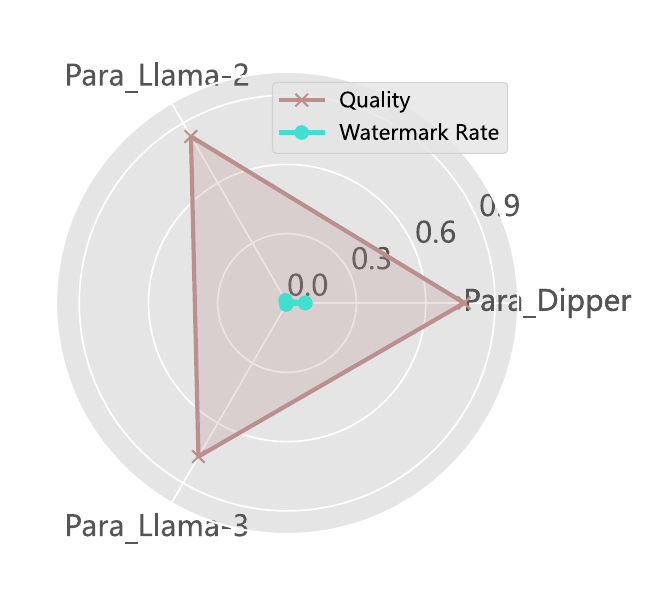}
\caption{Paraphrase}
\label{Paraphrase}
\end{subfigure}
\begin{subfigure}{0.39\columnwidth}
\includegraphics[width=\columnwidth]{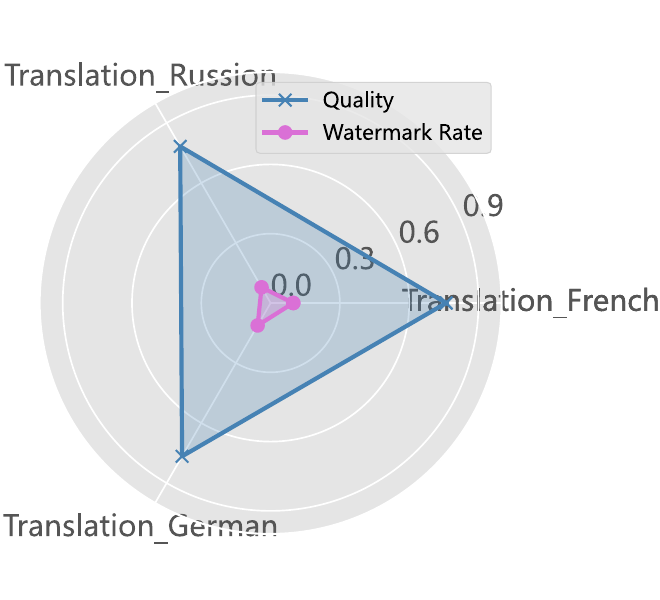}
\caption{Translation}
\label{Translation}
\end{subfigure}
\begin{subfigure}{0.39\columnwidth}
\includegraphics[width=\columnwidth]{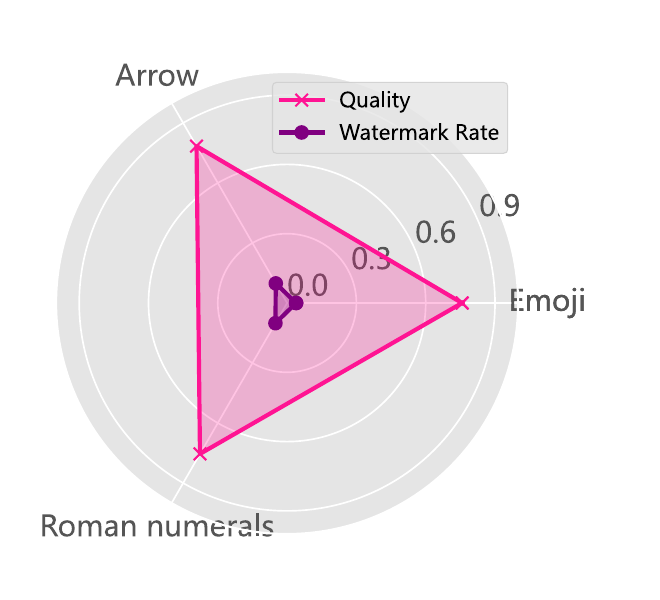}
\caption{Distill}
\label{Distill}
\end{subfigure}
\begin{subfigure}{0.39\columnwidth}
\includegraphics[width=\columnwidth]{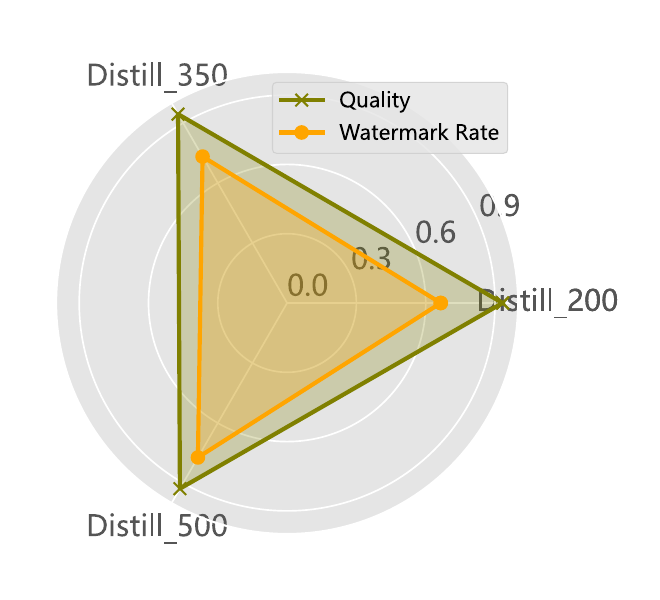}
\caption{Emoji}
\label{Emoji}
\end{subfigure}
\caption{Expanded parameters configuration of attacks. For \Cref{Helm}, the Lowercase, Expansion, and Contraction attacks are all belonging to the Helm~\cite{liang2022holistic} attack family.
Hence, we aggregate these attacks for collective analysis. Note that here we take KGW as the watermark as it performs the best in our previous evaluation.}
\label{fig:expanded attack}
\end{figure*}

\mypara{Quality \& Watermark rate}
Here we dive deeper into the quality and watermark rate performance of each watermark against different attacks.
The results are summarized in~\Cref{fig: results of watermark LLMs}.
Regarding the quality, we find that KGW, Unigram, Convert, Inverse, Exponential, and UniSpaCh preserve the quality to a large extent after different attacks, e.g., around 0.8.
However, for most of the watermarks, the watermark rate drops drastically to less than 0.6, and most of them are near 0.0, which means the watermark cannot be extracted after the attack.
This further emphasizes the vulnerability of the watermarks on texts under potential adversarial attacks.

\mypara{Impact of different attack methods}
From now on, focus on KGW to discuss the attacking effectiveness as KGW reaches the best performance in our previous evaluation.
We first discover that Typo, Token, Translation, and Emoji attacks are more effective than others.
For Typo, the attack has a significant impact on the model's output texts.
For Emoji and Token attacks, while these two attacks might only induce minor disruptions in the token generation process, they could potentially impact the distribution of the entire output token list, leading to a significant decrease in the watermark detection of KGW.
For Translation and Paraphrase attacks, both forms of attack entirely alter the output text of the model, while preserving the semantic integrity of the said text.
Hence, there is a great effect on the output text token which results in a reduction in the watermark rate.
Meanwhile, \Cref{Distill} demonstrates that the distillation attack is ineffective, as the distilled model retains a watermark rate exceeding 0.65.
Moreover, this watermark rate increases with the length of token sequences.
This is due to the fact that the distillation approach employed for the model does not inflict any damage on the model's output.

\mypara{Impact of attacking hyperparameters}
We then evaluate the watermark robustness with different hyperparameter settings for each attack.
The results are shown in \Cref{fig:expanded attack}.
We observe that, regarding different hyperparameter settings, the watermark quality is similar but the watermark rate may differ.
The empirical evidence gathered suggests that even slight alterations in these parameters can significantly impact the efficacy of the same attack, leading to substantial discrepancies in the outcomes.
For instance, the various forms of attack within the Helm attack family yield distinct outcomes contingent upon the specific type of attack employed.
Our hypothesis attributes this to the inherent disparities in the fundamental methodologies employed across various attacks.
Another two examples are the Modify attack and the Token attack.
Initially, the Modify attack exhibits a potent attack impact when the parameter is set to $(0.05,0.05,0)$.
However, the attack impact diminishes when the parameter is adjusted to $(0,0,0.05)$ and $(0,0,0.10)$.
This outcome can be deliberated upon from the perspective of the established definition of the Modify attack.
The attack corresponding to the initial parameter setting duplicates and removes certain words, while the attacks associated with the other two settings merely substitute some words.
It can be discerned that the deletion or addition of a word in a sentence can substantially influence the quality of the text.
Conversely, the impact is relatively minimal when a word is simply replaced with another.
Moreover, we find that the Token attack shows a more interesting phenomenon.
This attack bears a resemblance to the augmentation of the Modify attack within the scope of the token dimension.
Upon the deletion or insertion of random tokens into the output token list, there is a chance to substantially decrease the watermark rate.
However, the replacement of a few tokens at random positions is not stable and might only have a marginal effect on decreasing the watermark rate.

\mypara{Takeaways}
In summary, the extent of alteration in the output token sequence instigated by an attack dictates the magnitude of the watermark rate.
Looking ahead, we could advocate for the implementation of watermarks with more profound dimensions, such as the feature space of every token, as a countermeasure against those attacks.

\begin{figure*}[h]
    \centering
\begin{subfigure}{1.8\columnwidth}
\includegraphics[width=\columnwidth]{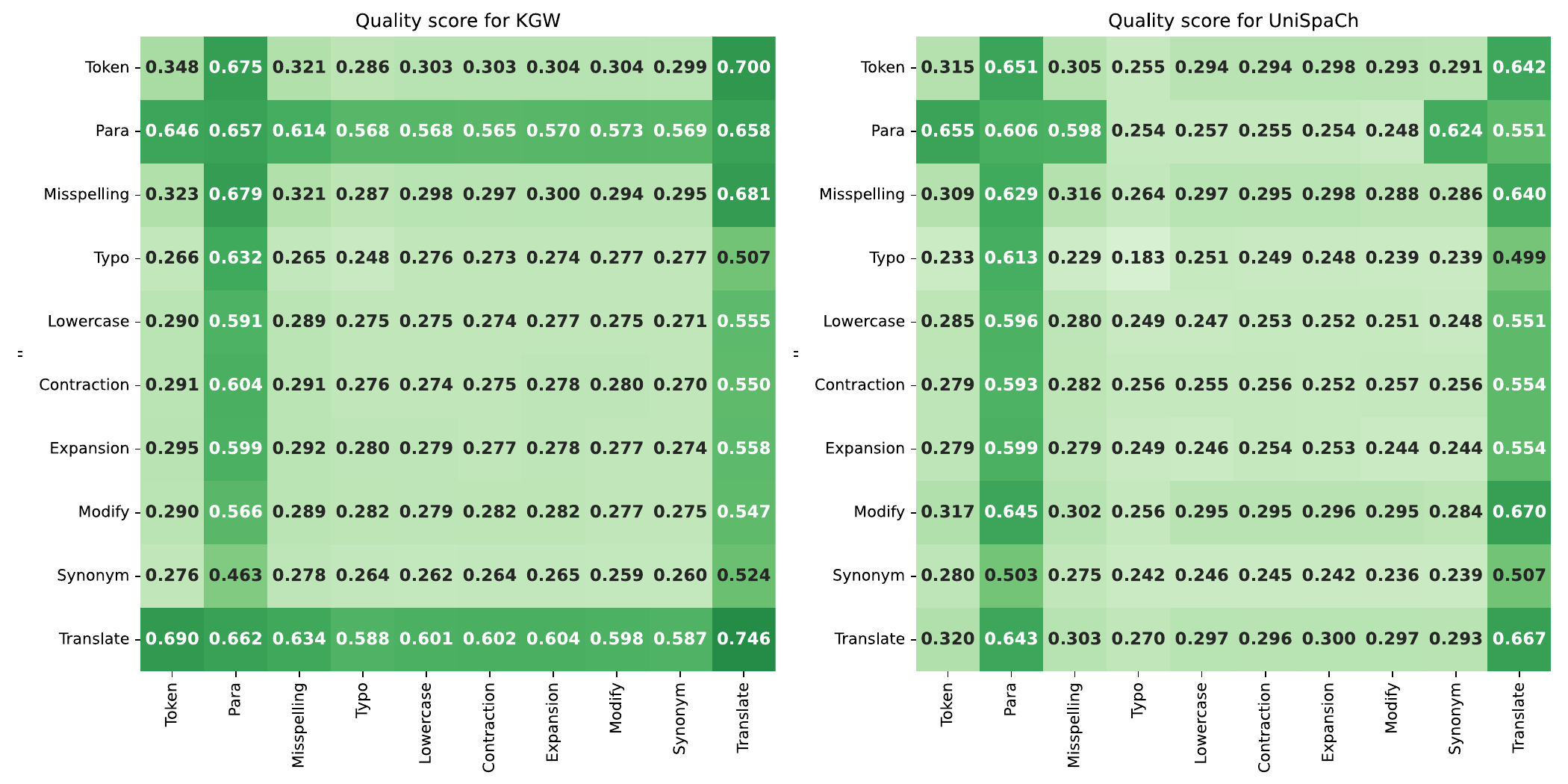}
    \caption{Quality score.}
    \label{fig:multi_quality}   
    \end{subfigure}
    \begin{subfigure}{1.8\columnwidth}
         \includegraphics[width=\columnwidth]{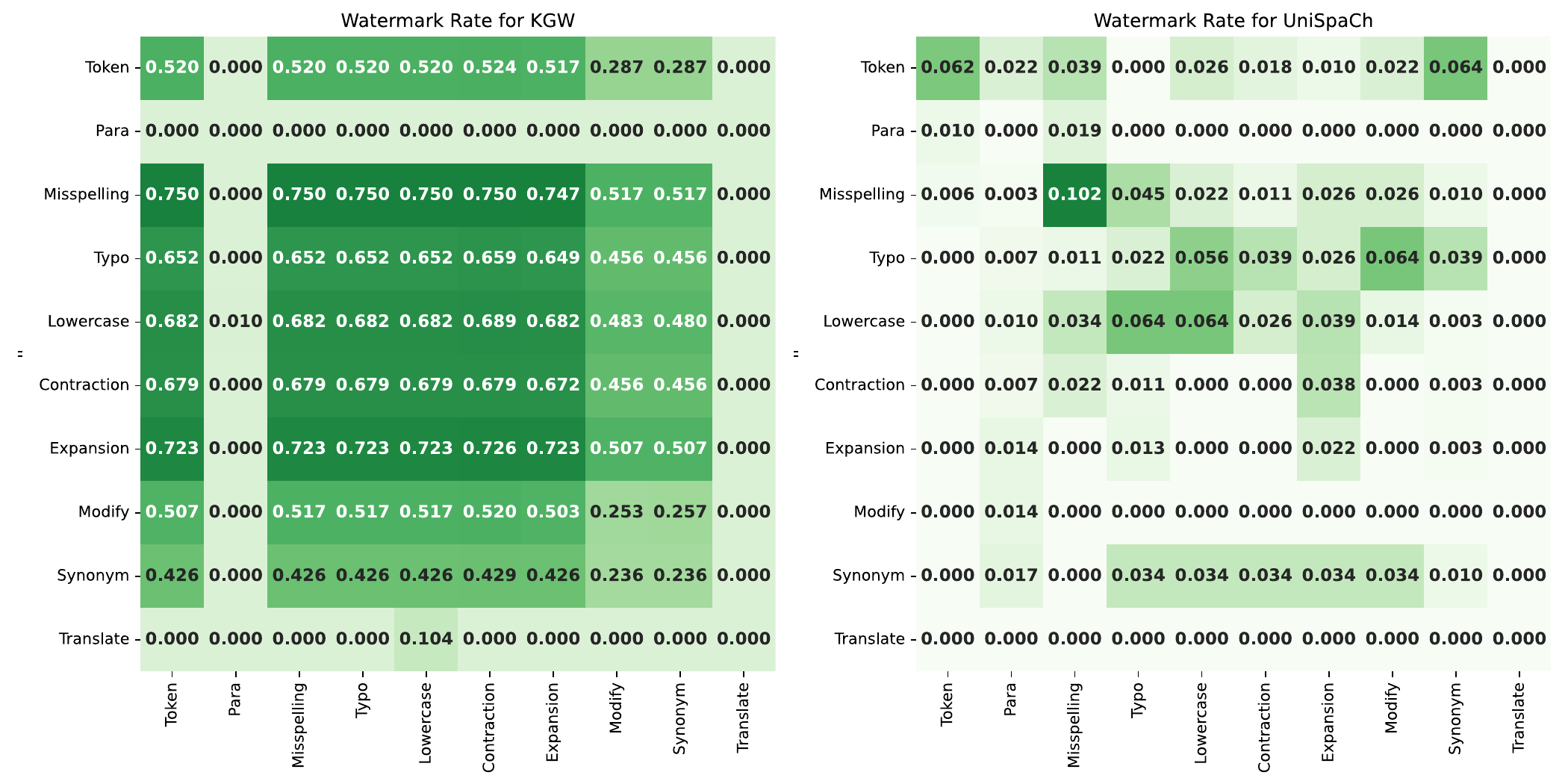}
    \caption{Watermark rate.}
    \label{fig: multi_detect}
    \end{subfigure}
    \caption{The quality score and watermark rate of KGW and UniSpaCh watermarking schemes after the combined attack.}
\end{figure*}

\subsection{Robustness Against Combined Attacks}

To more effectively illustrate the vulnerability of machine-generated text watermarks in real-world scenarios, we advocate for the implementation of combined attack strategies.
The optimal watermarks for pre-text (KGW) and post-text (UniSpaCh) are chosen as the target watermark algorithm in this experiment and the results are shown in \Cref{fig:multi_quality} and \Cref{fig: multi_detect}, respectively.
From each sub-figure, the rows represent the first attack and the columns represent the second attack.

\mypara{Results overview}
In general, KGW maintains a higher quality and watermark rate than UniSpaCh.
For instance, when applying Paraphrase first and Modify later, KGW reaches 0.5726 quality while UniSpaCh only has 0.2542 quality.
Meanwhile, the columns and rows representing Paraphrase and Translation attacks exhibit superior quality.
When Paraphrase and Translation are utilized as the secondary attack, it has been observed that this approach can even lead to an enhancement in the quality of the text.
However, the watermark rates are all reduced regardless of the attack order and attack category.

\mypara{Quality results analysis}
Recall that the quality score reflects whether the generated text can align with our real requirements and if the text remains consistent with the query's answer.
Hence, the text quality will be higher when the text is more informative and has fewer word errors.
When a generated text has been attacked by schemes such as Modify and Typo attacks, more grammatical errors will be generated in the attacked text, thereby causing quality degradation.
However, we can regard the Paraphrase and Translation attack as text correction tools that can be used to remove grammatical errors in the text.
This property of both attacks leads to an increase in text quality after the combined attack.

\mypara{Watermark rate results analysis}
Regarding the watermark rate, there is a great gap between the two watermarks we considered.
KGW has a higher watermark rate and most of the results range from 0.4 to 0.7.
Nevertheless, the watermark rate drops to 0 when the text is attacked by Paraphrase or Translation attacks. 
On the other hand, UniSpaCh's performance is poor and all results are close to 0.
We observe that the watermark rates are higher after the Paraphrase and Translation attack which converse with the results of KGW.

\mypara{KGW vs UniSpaCh}
Through comprehensive analysis of experimental results, we conclude that different watermark schemes exhibit distinct efficacy in withstanding combined attacks.
This is due to the inconsistency in the watermark verification methods.
For instance, the verification method of KGW calculates a secret key to divide the token's vocabulary into the green list and the red list and make a statistical analysis of the token percentage that belongs to the two lists.
Therefore, Paraphrase attacks and translation attacks both disrupt the token statistics by making significant changes to the text, rendering the watermark undetectable.
However, both two attacks cannot completely break the specific signals which is a watermark detection mark of UniSpaCh, which induces a slightly higher watermark rate (around 0.01).

\begin{table}[h]
\centering
\caption{Efficiency results of watermarking schemes (second).}
\begin{tabular}{llcc}
\toprule
\textbf{Category} & \textbf{Schemes} & \textbf{Injection} & \textbf{Detection} \\
\midrule 
\multirow{5}{*}{Pre-text }   & KGW& 4,032 & 364 \\
& Unigram& 3,888 & 146 \\
&Convert& 4,428 & 284 \\
&Inverse& 1,476 & 529 \\
&Exponential& 1,656 & 148 \\
\midrule
\multirow{3}{*}{Post-text } &  WHITEMARK & 84 & 4\\
& UniSpaCh & 126 & 6\\
& Linguistic& 113,472& 202,752\\
\bottomrule
\end{tabular}
\label{tab: injection and detection time}
\end{table}

\subsection{Efficiency}

Besides robustness, efficiency is also an important factor in evaluating the performance of watermarking schemes and removal attacks.
To this end, we further measure the runtime of watermarking schemes (including both the injection and detection process) and removal attacks by using the whole watermark generation dataset $\mathcal{D}_{\rm Gen}$.
Our implementation environment is on one NVIDIA A800 GPU.

\mypara{Watermark efficiency}
We summarize the runtime evaluation results of watermarking schemes in~\Cref{tab: injection and detection time}.
We could get the following observations:
(1) For the pre-text watermarking schemes, we notice that the watermark injection runtimes are roughly around 0.5 to 1 hour and the detection runtimes are all less than 600 seconds.
Taking into account the duration of text generation, the time allocated to watermarking is comparatively minimal.
This observation underscores that the operational efficiency of the watermark LLM falls within a tolerable threshold, making it available for real-world scenarios.
(2) For the post-text watermarks, we find that there is a great time gap among them.
For instance, the experiment results of WHITEMARK and UniSpaCh are both less than 150 seconds.
However, the Linguistic watermark is extremely slow with an injection/detection speed of 31.52/56.34 hours.
This empirical evidence underscores the significance of the efficiency of injection and detection for watermarking in the overall evaluation of the watermark’s performance.
Therefore, we can conclude that even when a watermark exhibits high robustness if it is characterized by low efficiency, its applicability in real-world scenarios is limited.
This highlights the need to balance both efficiency and robustness in the watermark design.

\begin{figure*}[!t]
\centering
\begin{subfigure}{1\columnwidth}
\includegraphics[width=\columnwidth]{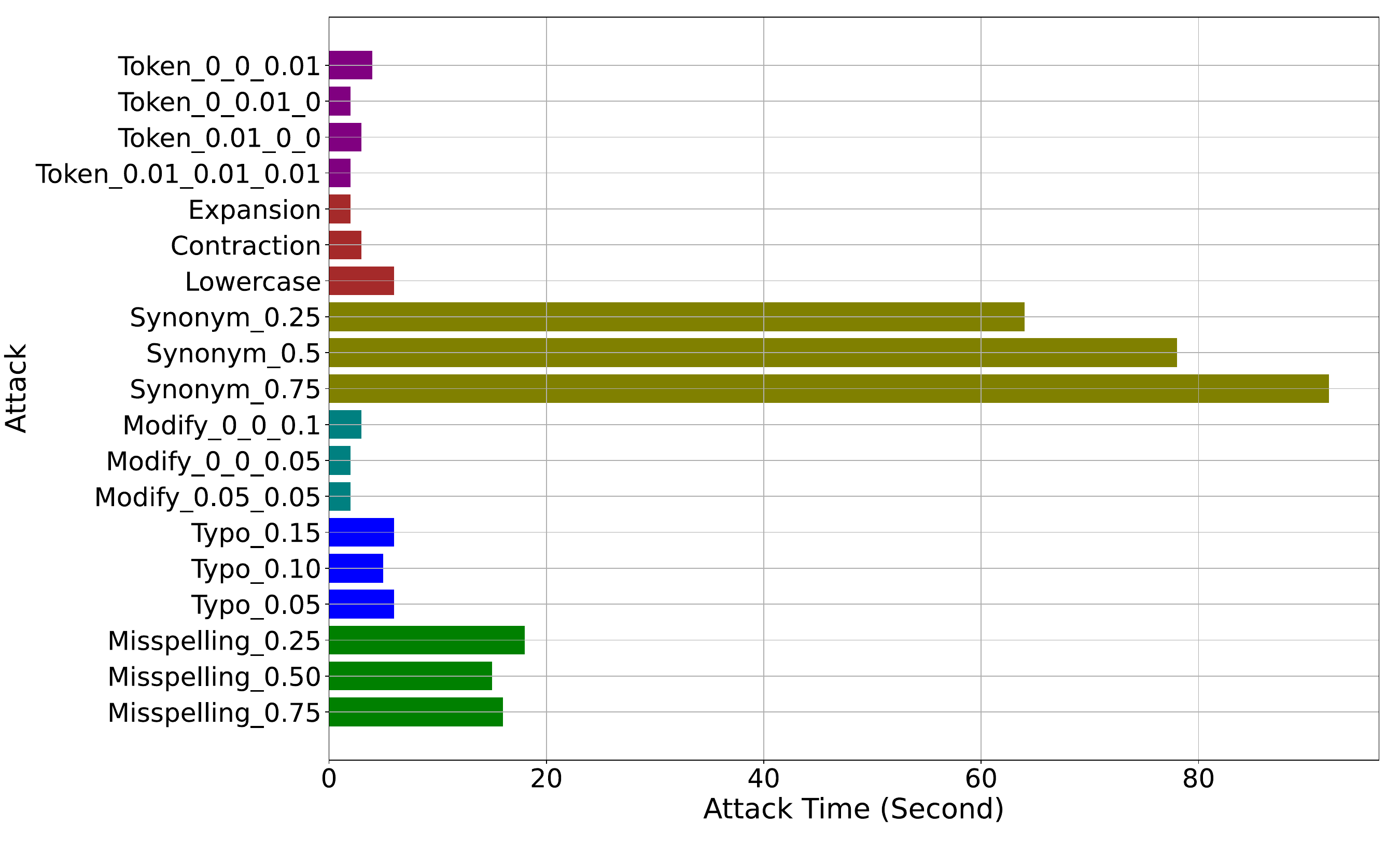}
\caption{Attacks with second-level runtime.}
\label{Fig: attack timea}
\end{subfigure}
\begin{subfigure}{1\columnwidth}
\includegraphics[width=\columnwidth]{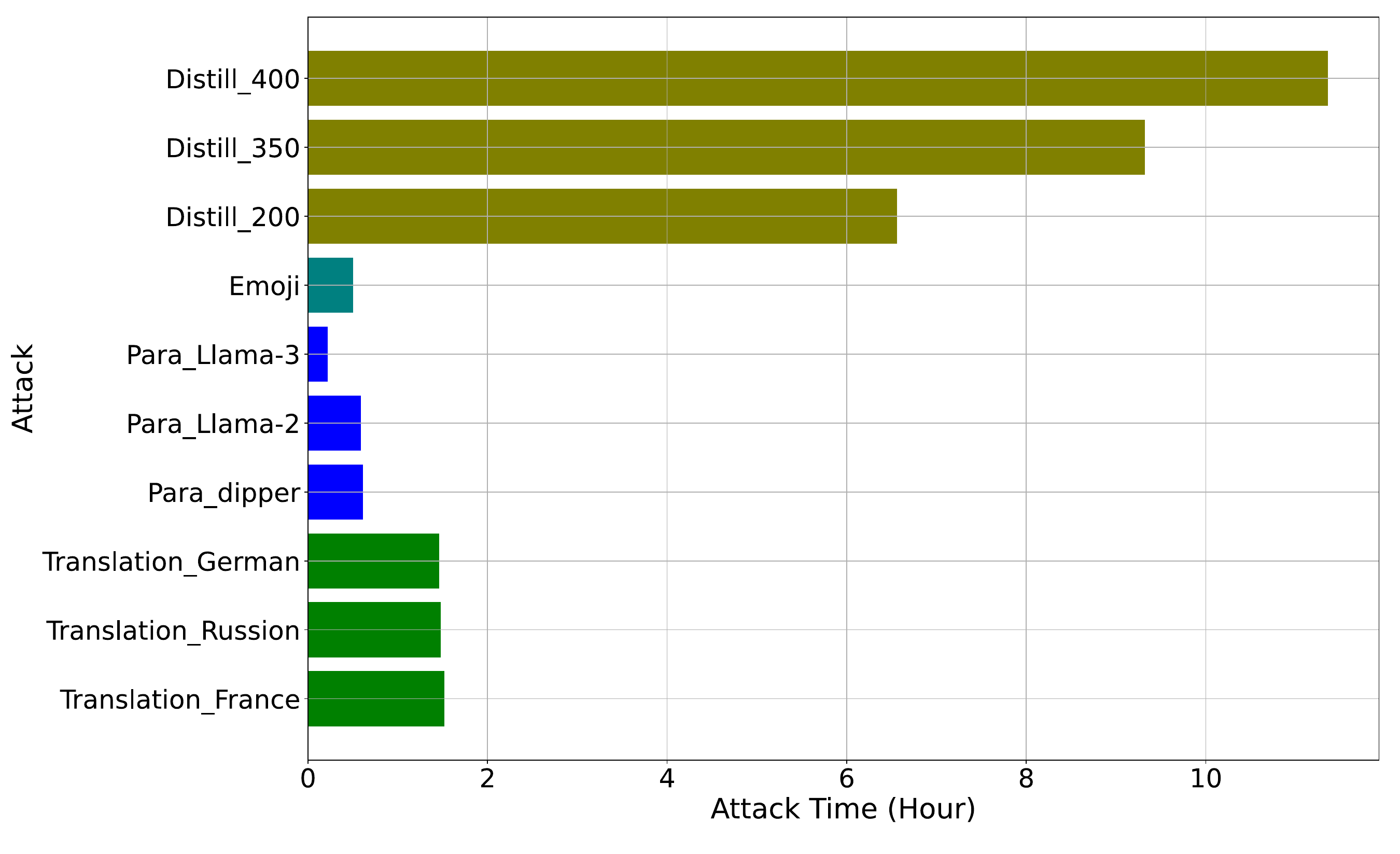}
\caption{Attacks with hour-level runtime.}
\label{Fig: attack timeb}
\end{subfigure}
\caption{Efficiency results of watermark removal attacks.
Here we fix the watermarking scheme to KGW.}
\label{fig: results of watermark text}
\end{figure*}

\mypara{Attack efficiency}
We summarize the runtime for different watermarking removal attacks with different hyperparameters in \Cref{fig: results of watermark text}. 
We observe that, while Paraphrase and Translation attacks demonstrate high effectiveness (\Cref{fig:expanded attack}), their efficiency is relatively low (\Cref{Fig: attack timeb}). 
In contrast, Typo and Token attacks exhibit better performance in both effectiveness and efficiency, suggesting they are more optimal attack choices.

\subsection{Imperceptibility}

\mypara{Evaluation methodology}
As pointed out in~\cite{lukas2021sok,piet2023mark}, an ideal watermarking scheme should exhibit good imperceptibility. 
In other words, imperceptibility measures how well the watermarking information is hidden such that the watermarking scheme doesn't affect the readability, coherence, or naturalness of the AI-generated texts.
To quantify imperceptibility, we leverage LLM to judge if a text is an AI-generated text that contains a watermark and uses the classification accuracy as the imperceptibility results.
As the texts to be judged are all watermarked texts, consequently, a lower imperceptibility value indicates that LLM cannot recognize the hidden watermark information.
Here we regard Llama-3-8B-instruct~\cite{llama3modelcard} as the classifier and use the following prompt to classify the texts.

\begin{tcolorbox}[colback=gray!25!white, size=title,boxsep=1mm,colframe=white, after={\vskip0mm}]
\tiny
\texttt{<|begin\_of\_text|><|start\_header\_id|>\textbf{system}<|end\_header\_id|>}

\textit{You are an assistant model to help me judge if the given text has a watermark and is generated by a machine, and please output your judgment 'True' or 'False' without any other content after it.}

\textit{\#\#\# Text: \{text\}}

\texttt{<|eot\_id|><|start\_header\_id|>\textbf{user}<|end\_header\_id|>}

\textit{Please provide the judge result.}

\textit{\#\#\# Result:} 

\texttt{<|eot\_id|><|start\_header\_id|>\textbf{assistant}<|end\_header\_id|>}

\end{tcolorbox}

\begin{figure}[t]
    \includegraphics[width = 1\columnwidth]{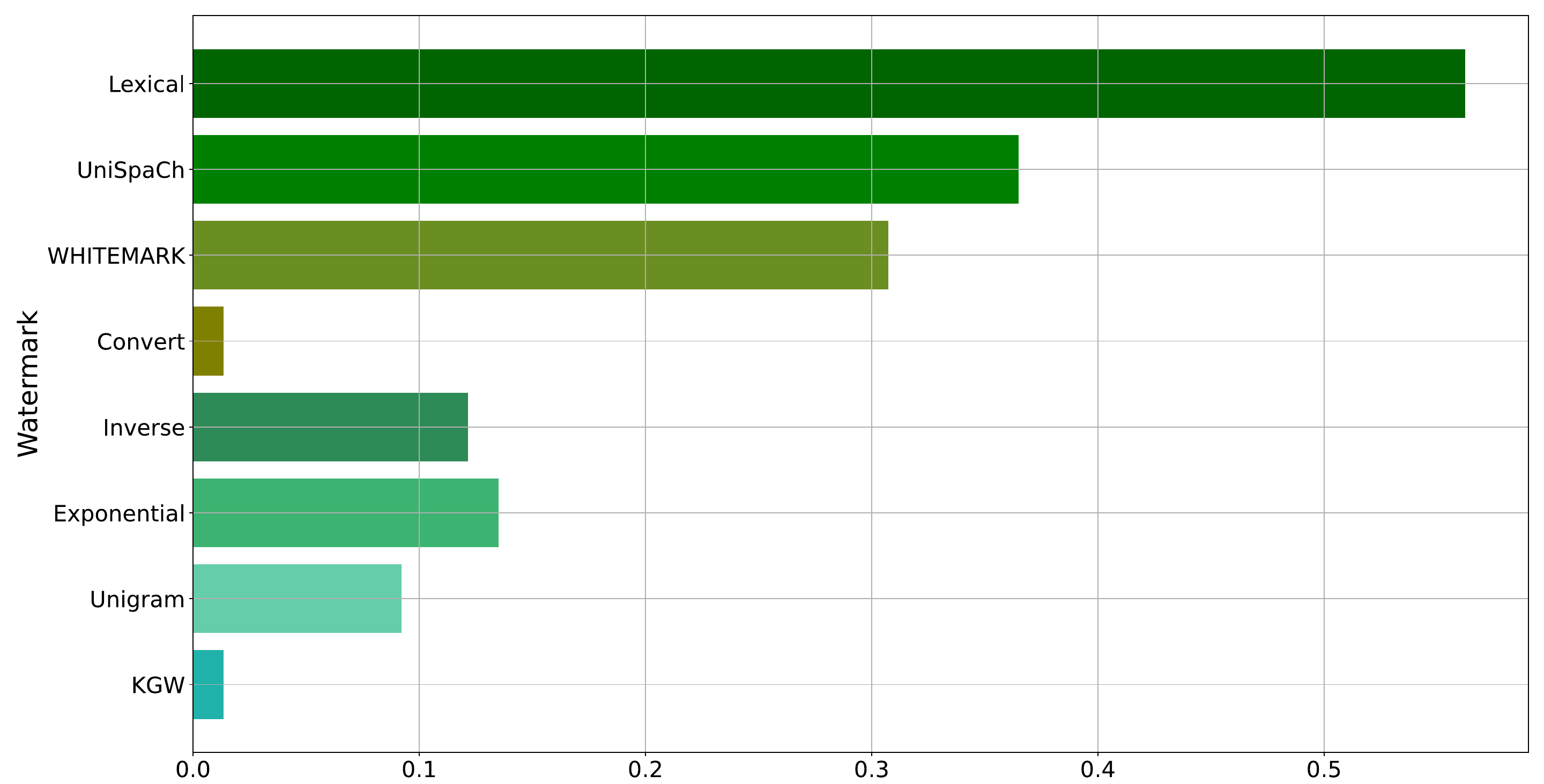}
    \caption{Imperceptibility results ($\downarrow$) of different watermarking schemes.}
    \label{Imperceptibility}
\end{figure}

\mypara{Evaluation results}
We evaluate the imperceptibility of each watermarking scheme and present the results in \Cref{Imperceptibility}.
We observe that pre-text watermarks achieve better imperceptibility (lower value is better).
This is because the pre-text watermarking schemes just modify the logits or the token-sampling strategy during the inference process, thus, there is no obvious watermark signal left in the text that the judgment LLM can capture.
In contrast, all three post-text watermarking schemes need to embed signals into the watermarked text, thereby causing a worse imperceptibility.
For instance, both WHITEMARK~\cite{sato2023embarrassingly} and UniSpaCh~\cite{UniSpaCh} schemes inject the specific whitespace characters into the text to inject the watermark.
Lexical watermark~\cite{yang2023watermarking} will replace the origin word with the new unusual word which may be viewed as a signal word by the assistant model.
The degradation in the performance of imperceptibility is attributed to these particular signals.

\mypara{Takeways}
It can be inferred that the practice of injection watermarks by altering the output text extensively is sub-optimal and the pre-text watermarking schemes have great potential to be used in real-world scenarios.

\section{Related Work}

\subsection{Machine-Generated Text Watermark}

Watermark of machine-generated text has been extensively studied~\cite{8268096,Rizzo17,articlepre}.
It can be viewed as a variant of steganography~\cite{10.1007/11551492_2,math9212829} with one-bit or multi-bit messages.
The text watermark has a strong requirement that the watermarked text and unwatermarked text are indistinguishable.
These watermark schemes can be broadly divided into two categories: pre-text watermarks and post-text watermarks.
The pre-text watermarks inject the watermark into texts before its generation which can occur during the training time~\cite{li2023watermarking,peng-etal-2023-copying,tang2023did,li2023protecting}, modify the logits~\cite{kirchenbauer2023watermark,zhao2023provable,lee2024wrote,hu2023unbiased,fernandez2023bricks,fairoze2024publiclydetectable,kirchenbauer2024reliability,liu2024unforgeable,ren2024robust} and modify the token sampling strategy~\cite{cryptoeprint:2023/763,kuditipudi2023robust,aaronson2022watermarking} during the inference time.
The Post-text watermarks inject the watermark into the generated texts, focusing on format-level~\cite{sato2023embarrassingly,UniSpaCh} or lexical-level~\cite{yang2023watermarking,zhang2023remark} modifications.

In this work, we mainly focus on watermarks against generated texts.
Note that recent works also develop watermarks against model weights~\cite{ABCPK18,RCK18,CHZ22,li2023watermarking}, which focus on protecting the intellectual property of the model.
Cong et al.~\cite{CHZ22} develop SSLGuard, which is a watermarking scheme against self-supervised encoders.
Li et al.~\cite{li2023watermarking} propose a watermarking scheme called weight quantization that modifies the model weight based on fine-tuning.
This method embeds the watermark in the gap between full-precision weights and quantized weights.

Another line of works focuses on detecting whether the text is real or generated by models by training a classifier or setting 
 threshold for pre-defined metrics~\cite{SBCAHWRW19,GSR19,MLKMF23,GZWJNDYW23,MLKMF23,HSCBZ24}.
Mitchell et al.~\cite{MLKMF23} develop DetectGPT, which leverages multiple random perturbations to the text and measures the change of the model's log-likelihood.
He et al.~\cite{HSCBZ24} propose MGTBench, which is a benchmarking framework including different types of detection methods.

\subsection{Watermark Removal Attacks}

To evade the watermark detection on texts, researchers have developed various methods, which can be categorized into pre-text attacks and post-text attacks as well.

For pre-text attacks, Kirchenbauer et al.~\cite{kirchenbauer2023watermark} introduce an attack designed to undermine the generative capabilities of LLMs by prompting them to alter their output into a desired format.
Similarly, Goodside\cite{goodside2023adversarial} proposes the ``emoji attack'', where the LLM is instructed to insert an emoji after each word, which is subsequently removed.
This method disrupts the token sequence, effectively eliminating any embedded watermarks.
Gu et al.~\cite{gu2024learnability} aims to train a student model to emulate the behavior of the target model.
Concretely, it first generates a watermarked text dataset from the target model and then fine-tunes the student model on this dataset using the standard language modeling cross-entropy loss.

Regarding post-text attacks, Liang et al.~\cite{liang2022holistic} develop HELM, which contains various attacks such as Contraction, Expansion, Lowercase, Misspelling, and Typo.
Piet et al.~\cite{piet2023mark} also consider Synonym attacks which leverage WordNET to find candidate words and replace them with synonyms by querying GPT-3.5.
Krishana et al.~\cite{krishna2023paraphrasing} develop dipper, which is a paraphrase generation model (11B parameters) to paraphrase texts.
Kuditipudi et al.~\cite{kuditipudi2023robust} propose the Token attack, a novel attack that modifies output texts at the token level. In this attack, the output texts are re-encoded into token lists. Tokens are then randomly sampled from these lists according to a specified distribution. The sampled tokens are subjected to operations such as replacement, deletion, or iterative insertion.

\section{Conclusion}

In this paper, we systematically categorize watermarks and attacks on machine-generated texts into pre-text and post-text classes, providing a structured framework for evaluating their performance.
Our comprehensive assessment involved eight watermarks and twelve attacks, resulting in 87 possible scenarios, allowing us to rigorously test the robustness, efficiency, and imperceptibility of existing watermarking techniques.

Our findings indicate that KGW and Exponential watermarks are currently the most effective, offering good text quality and relatively high watermark retention rates after various attacks.
However, despite their relative effectiveness, these watermarks remain vulnerable to a variety of attacks.
The watermark rate for KGW, for example, significantly drops under certain attack combinations, highlighting the ongoing challenges in watermark resilience.

Efficiency analysis revealed that both KGW and Exponential are efficient in watermark injection and detection.
In terms of imperceptibility, pre-text watermarks perform better as they are embedded within token distributions and do not disrupt text readability, unlike post-text watermarks which are more detectable due to their token manipulations.
Regarding attacks, post-text attacks generally prove more efficient and practical compared to pre-text attacks, as they do not require modifications to the model's weights.
Our study also demonstrates that adversaries can significantly enhance the effectiveness of their attacks by combining different methods. This underscores the urgent need for developing more robust watermarking solutions capable of withstanding diverse and combined attack strategies.

In conclusion, while our work highlights the promise of watermarking techniques in safeguarding the authenticity of machine-generated texts, we also expose the limitations and vulnerabilities of current approaches.
Our research emphasizes the necessity for continued innovation and development of more resilient watermarking schemes to ensure the integrity and reliability of digital communications.
Our code and data will be made publicly available to support further research in this domain.

\newpage
\bibliographystyle{plain}
\bibliography{generated,additional}

\end{document}